\documentclass[journal,12pt,draftclsnofoot,onecolumn]{IEEEtran}

\usepackage{etex}

\usepackage{cite}
\usepackage{tikz}
\usetikzlibrary{arrows,positioning,decorations.markings}

\usepackage{verbatim}
\ifCLASSINFOpdf

\else

\fi
\usepackage{array}
\usepackage{algorithm}
\usepackage{algcompatible}
\usepackage{algpseudocode}
\usepackage{float}
\usepackage{array}
\usepackage{psfrag}
\usepackage{balance,multirow}
\usepackage[hidelinks]{hyperref}
\usepackage{caption}
\usepackage{booktabs}\usepackage[printonlyused,withpage]{acronym}
\usepackage{subcaption}
\usepackage{mathtools}
\usepackage{breakurl}
\usepackage{pstricks, pst-node, pst-plot, pst-circ}
\usepackage{moredefs}
\usepackage{graphicx}
\usepackage{pgfplots}

\usepackage{setspace}

\usetikzlibrary{positioning}
\usetikzlibrary{arrows}
\usepackage{acronym}

\usepackage{optidef}
\usepackage{dsfont}
\usepackage{amssymb}
\usepackage{amsmath}
\usepackage{amssymb}
\usepackage{kbordermatrix}

\newcommand{\norm}[1]{\left\lVert#1\right\rVert}

\usepackage{stackengine}
\def\delequal{\mathrel{\ensurestackMath{\stackon[1pt]{=}{\scriptstyle\Delta}}}}

\def\ve#1{{\mathchoice{\mbox{\boldmath$\displaystyle #1$}}%
{\mbox{\boldmath$\textstyle #1$}}%
{\mbox{\boldmath$\scriptstyle #1$}}%
{\mbox{\boldmath$\scriptscriptstyle #1$}}}}

\usepackage[utf8]{inputenc}
\usepackage{chngcntr}
\usepackage{apptools}
\AtAppendix{\counterwithin{lemma}{section}}
\AtAppendix{\counterwithin{proposition}{section}}
\newtheorem{proposition}{Proposition}
\newtheorem{lemma}{Lemma}
\newtheorem{corollary}{Corollary}
\newtheorem{theorem}{Theorem}
\newtheorem{definition}{Definition}

\newtheorem{notation}{Notation}

\usepackage{chngcntr}
\begin{document}

\title{Full Characterization of Optimal Uncoded Placement for the Structured Clique Cover Delivery of Nonuniform Demands}

\author{Seyed Ali Saberali, Lutz Lampe and Ian Blake}

\maketitle
\begin{abstract}
We investigate the problem of coded caching for nonuniform demands when the structured clique cover algorithm proposed by Maddah-Ali and Niesen for decentralized caching is used for delivery. We apply this algorithm to all user demands regardless of their request probabilities. This allows for coding among the files that have different request probabilities but makes the allocation of memory to different files challenging during the content placement phase. As our main contribution, we analytically characterize the optimal placement strategy that minimizes the expected delivery rate under a storage capacity constraint.  It is shown that the optimal placement follows either a two or a three group strategy, where a set of less popular files are not cached at all and the files within each of the other sets are allocated identical amounts of storage as if they had the same request probabilities. We show that for a finite set of storage capacities, that we call the base-cases of the problem, the two group strategy is always optimal.   For other storage capacities, optimal placement is achieved by memory sharing between certain base-cases and the resulting placement either follows a two or a three group strategy depending on the corresponding base-cases used. We derive a polynomial time algorithm that determines the base-cases of the problem given the number of caches and popularity distribution of files. Given the base-cases of the problem, the optimal memory allocation parameters for any storage capacity are derived analytically.

\end{abstract}
\newpage
\doublespacing
\section{Introduction}\label{sec:intro}
\subsection{Background}
The next generation wireless communication networks deploy a dense composition of short-range and low-power small-cells and combine them with the macrocells into heterogeneous networks. This architecture promotes localized communications and effectively increases the area spectral efficiency of the network. The performance of such networks is however challenged by the congestion of the backhaul links that connect the small-cells to the backbone communications network during the peak traffic hours.
Caching at the edge is a promising technique to alleviate the backhaul congestion through the storage of popular content closer to the end users \cite{Liu:2016,Zeydan:2016,Zhao:2016,Golrezaei:2013,Kiskani:2017,Pappas:2018,Song:2017,Maddah_Magazine:2016}.

Coded caching \cite{Maddah_Magazine:2016,Maddah_limits:2014,Maddah_decentralized:2014} is a novel approach for content caching in  a network that consists of multiple caching nodes which communicate with a central server over a shared broadcast channel. This technique benefits from network coding and coded multicasting to gain superlinear reduction in the data delivery load on the shared link as the cache capacity increases. In particular, during a placement phase, popular content is carefully distributed over the different storage nodes such as to create coding opportunities among the caches. During a delivery phase, the content that is requested but is missing from the caching nodes is delivered to them by the central server's transmissions over the shared link. The server exploits the coding opportunities created during placement to embed the missing content requested by multiple caches in a single message that every target cache can decode for its desired content. The load on the shared link is referred to as the delivery rate. 

The coded caching proposed by Maddah-Ali and Niesen in their seminal work \cite{Maddah_decentralized:2014} efficiently utilizes the side information that each cache has about the requests of the other caches in order to build server's coded messages. This delivery algorithm can be viewed as a structured way of clique-covering  the vertices of the side information graph that characterizes the availability of the content requested by each cache in the other caches. As a result, we call this delivery algorithm \acrodef{scc}[SCC]{Structured Clique Cover} \ac{scc} procedure throughout this paper.  

The seminal works \cite{Maddah_limits:2014} and \cite{Maddah_decentralized:2014} aimed at minimizing the peak delivery rate when different files in the library are equally likely to be requested by users, i.e., when  the user demands are uniform. However, a more practical scenario concerns caching of files with different popularity levels. In this scenario, it is expected  to allocate more storage to the caching of the more popular files during placement. This idea is followed in several works in the literature \cite{Maddah_nonuniform:2014,Digavi:2015,Caire:2014,Zhang:2015,Caire:2017,Yu:2017}. 
\vspace*{-.4cm}
\subsection{Related Work}\label{sec:related}
Two major approaches are followed for coded caching with nonuniform demands. The first approach is based on the grouping of files into different popularity groups based on their request probabilities \cite{Maddah_nonuniform:2014,Digavi:2015,Zhang:2015}. With the files in each group having relatively similar request probabilities, the \ac{scc} algorithm is applied separately to the requests belonging to each group for delivery. The advantage of this method is the simplicity of the analysis of the expected rate. Its main disadvantage is that it limits the utilization of coding opportunities to the files that are within each popularity group. The  design objective in this approach is to find the grouping of files that achieves the lowest expected rate. A higher number of groups provides higher degrees of freedom to assign different amounts of storage to files with different request probabilities. On the other hand, the larger the number of groups is, the more underutilized are the coding opportunities among the different groups.
In \cite{Zhang:2015}, the library is grouped into two categories of popular and unpopular files. The requests for popular files are delivered by the \ac{scc} algorithm while the requests of unpopular files are delivered through  uncoded messages. This is an extreme  case of the grouping approach and its expected rate is shown to be at most a constant factor away from the information theoretic lower bound, independent of the file popularity distribution.

The second approach is followed in \cite{Caire:2017} and \cite{Yu:2017} which applies the \ac{scc} algorithm to all the user demands regardless of their request probabilities and the amount of storage allocated to each file. For any fixed placement of content, this delivery scheme outperforms the previously discussed group-based delivery. However, optimization of the amount of memory allocated to each file is challenging because of the complicated interplay between the memory allocation parameters and the expected delivery rate. References \cite{Caire:2017} and \cite{Yu:2017} use a convex optimization formulation of the memory allocation problem which aims to minimize the expected delivery rate for a given storage capacity per cache. We refer to this problem as \acrodef{rmsc}[RMSC]{Rate Minimization with Storage Constraint} \ac{rmsc}.  Reference \cite{Yu:2017} has followed a numerical approach to solve the \ac{rmsc} problem and is mainly focused on reducing the computational complexity of the numerical analysis involved.  In contrast, \cite{Caire:2017} follows a theoretical approach to find structural properties in the optimal solution of the problem.
\vspace*{-.3cm}
\subsection{Our Contributions}\label{subsec:contrib}
The results provided in \cite{Caire:2017} do not capture specific properties of the optimal solution which can considerably facilitate solving the memory allocation problem. In this work, we find such structures in the optimal solution and solve the \ac{rmsc} problem analytically when user demands are nonuniform and the \ac{scc} procedure is used for delivery. In particular, we will show that such properties enable the derivation of the optimal solution based on a search over a finite set of points. The cardinality of this set scales linearly with the product of the number of caches and the number of files, which is considerably small compared to the continuous search space of the optimization problem in \cite{Caire:2017}. The properties that we derive also provide a unifying interpretation of the optimal placement strategy for both uniform and nonuniform popularity distribution of files, as we will discuss in the remainder of this section.

As the first structural property, we show that for instances of the problem with cache capacities that belong to a finite set $\mathcal{M}$, the optimal placement for \ac{rmsc} follows the simple pattern of splitting the library of files into two groups. One group consists of less popular files and the files in this group are not cached at all. The files in the second group are cached but are treated as if they had the same request probabilities. We call such instances of \ac{rmsc} the \textit{base-cases}.

For instances of the problem that are not among the base-cases, we prove that the optimal solution is achieved by a convex combination of the solutions to certain base-cases of the problem. This solution is identical to the placement parameters obtained by memory sharing between the two base-cases of the \ac{rmsc} problem. Memory sharing is already shown to be optimal when demands are uniform  \cite[Lemma~5]{Caire:2017}, \cite[Theorem~1]{Yu:2017}, \cite[Proposition~1]{ours:arxiv}. Hence, this result shows that memory sharing is also optimal for nonuniform demands when applied to the appropriately chosen instances of the problem. To elaborate, let $K$, $N$ and $M$ be the number of caches, files in the library and files that each cache can store, respectively.  For optimal placement of identically popular files when \ac{scc} delivery is used, we have the following \cite{ours:arxiv,Caire:2017,Yu:2017}:
\begin{itemize}
\item All files are treated identically during placement, in particular, the same amount of storage is allocated to the caching of each file.
\item For a cache size $M$ that corresponds to an integer value of $t=K\frac{M}{N}$, the optimal placement breaks each file into $\binom{K}{t}$ nonoverlapping segments. Then, it exclusively stores each one of the segments in exactly one of the $\binom{K}{t}$ subsets of caches that have cardinality $t$. We refer to these cases of the problem as the base cases and denote by $\mathcal{M}$ the set of corresponding cache sizes $\{\frac{1}{K}N,\frac{2}{K}N,\ldots,N\}$.
\item For other cache capacities, the optimal placement can be obtained by memory sharing between the optimal placements for two instances of the problem with cache capacities $M_l=\max\{m\in\mathcal{M}\mid m<M\}$ and $M_u=\min\{m\in\mathcal{M}\mid M<m\}$.\footnote{The idea of memory sharing for uniform demands was presented in \cite{Maddah_limits:2014} as an achievable scheme when $t$ is not an integer. References \cite{ours:arxiv, Caire:2017,Yu:2017} proved that memory sharing is optimal for \ac{scc} delivery when demands are uniform.}
\end{itemize} 
We prove that a similar pattern exists in the optimal placement for nonuniform demands. In particular, we propose an algorithm with worst-case complexity of $O(K^2N^2)$ to derive the set $\mathcal{M}$ given a nonuniform popularity distribution for the files. If $M\not\in\mathcal{M}$, the optimal placement is obtained by memory sharing between $M_l,M_u\in\mathcal{M}$ as it was done for uniform demands  using the derived set $\mathcal{M}$. In this case, optimal placement either follows a two or a three group strategy depending on the specifics of the two corresponding base-cases used. 

As a result of our analysis, we notice that the memory allocated to different files does not show a gradual and smooth increase as the request probability increases. Instead, for base-cases where the two-group strategy is optimal, the memory allocation exhibits a binary behavior, i.e.,  as the request probability increases the amount of memory allocated to the files shows an abrupt increase from zero to a new level at a certain request probability and remains at that level thereafter. A similar trend exists for non base-cases, but there might be two thresholds on request probabilities where the jumps in the memory allocated to files occur.

The remainder of this paper is organized as follows. The setup of the problem and formulation of the expected rate and the storage used in terms of placement parameters are presented in Section~\ref{sec:overview}. The \ac{rmsc} problem is formulated in Section~\ref{sec:formulation} and a duality framework is proposed for it. Structures in the optimal solution of \ac{rmsc} for the base-cases are derived in Section~\ref{sec:jrsm}. In Section~\ref{sec:analysis_rmsc}, we propose an algorithm to identify the base-cases of the \ac{rmsc} problem for any given popularity distribution of files and  we derive the optimal solution of \ac{rmsc}. We conclude the paper in Section~\ref{sec:con}.

\section{Problem Setup and Formulation of Utilized Storage and Expceted Delivery rate}\label{sec:overview}
\subsection{Problem Setup}
\tikzset{
  treenode/.style = {align=center, inner sep=0pt,text centered },
  arn_r/.style = {treenode, draw=none, rectangle,rounded corners=1mm,
    fill=white, text width=4.7em,text height=1.7em,text depth=4em},
   arn_r_ser/.style = {treenode, draw=none, rectangle,rounded corners=1mm,
    fill=white, text width=4.7em,text height=1.6em,text depth=5em},
  arn_c/.style = {treenode, circle, black,draw=none,
    fill=white, text width=2em},
  arn_x/.style = {treenode, rectangle, draw=none,
    minimum width=0.0em, minimum height=0.0em}
}
\begin{figure}
\centering
\begin{tikzpicture}[>=stealth',level/.style={sibling distance = 4cm/#1,
  level distance = 1.5cm}] 
\node [arn_r_ser] {Server\\\medskip\includegraphics[scale=.15]{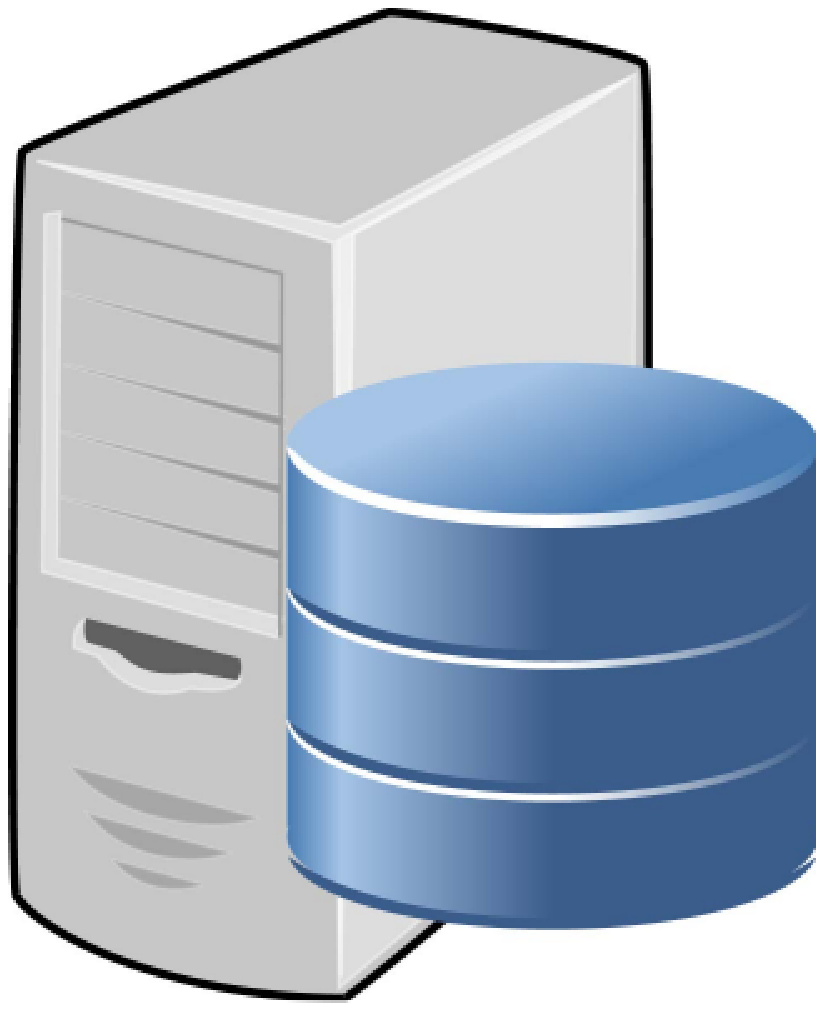}}
    child{ node [arn_x] {}[level/.style={sibling distance = 2cm,level distance = .7cm}] 
            child{ node [arn_x]{} [level/.style={level distance = 1.5cm}] 
            		child{node [arn_r] {\small{Cache $1$}\\\medskip
\includegraphics[scale=0.65]{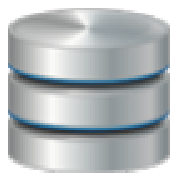}
            		} [level/.style={level distance = 2.2cm},->] 
            			child{node[arn_c] {\includegraphics[scale=0.14]{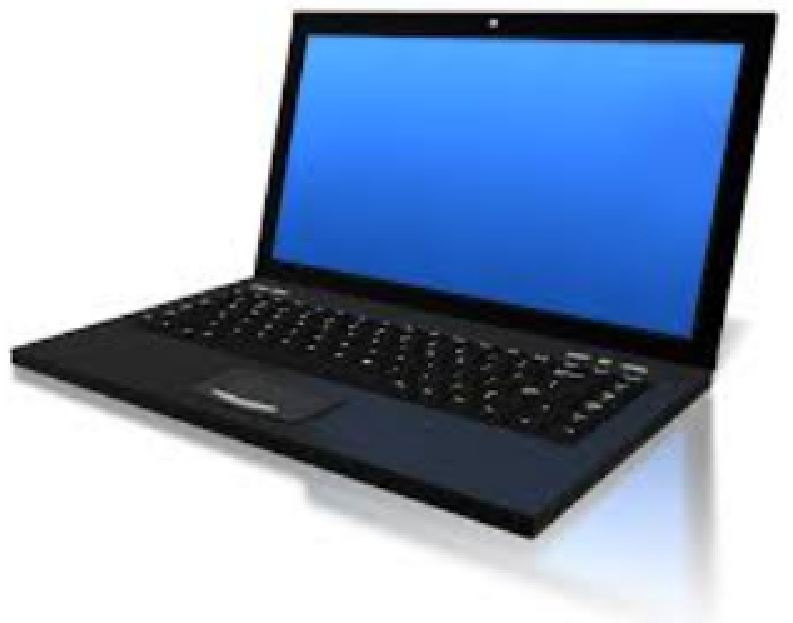}}
            			}
            		}
            }
            child{ node [arn_x]{} [level/.style={level distance = 1.5cm}]
            		child{node [arn_r] (a) {\small{Cache $2$}\\\medskip
\includegraphics[scale=0.65]{database.eps}
            		}[level/.style={level distance = 2.2cm},->] 
            			child{node[arn_c] (aa) {\includegraphics[scale=0.14]{laptop.eps}} 
            			}
            		}
            }
            child{ node [arn_x]{} [level/.style={level distance = 1.5cm}]
            		child{node [arn_r] (b) {\small{Cache $K\!-\!1$}\\\medskip
\includegraphics[scale=0.65]{database.eps}
            		}[level/.style={level distance = 2.2cm},->]  
            			child{node[arn_c] (bb) {\includegraphics[scale=0.14]{laptop.eps}} 
            			}
            		}
            }
            child{ node [arn_x]{} [level/.style={level distance = 1.5cm}]
            		child{node [arn_r]{\small{Cache $K$}\\\medskip
\includegraphics[scale=0.65]{database.eps}
            		}[level/.style={level distance = 2.2cm},->]  
            			child{node[arn_c] {\includegraphics[scale=0.14]{laptop.eps}} 
            			}
            		}
            }                                          
        }
;        
\path (a) -- (b) node [midway] {$\cdots$};      
\end{tikzpicture}
\vspace*{.1cm}
\caption{{A network with $K$ caches and a central server. }
}\label{config_fig}
\end{figure}
We consider the canonical network of multiple caches and a central server as modeled in \cite{Maddah_nonuniform:2014} for the general case where user demands can be nonuniform (see Fig.~\ref{config_fig}). In particular, we consider a library of $N$ files, each of length $F$ bits and a network of $K$ caches, each with storage capacity of $MF$ bits. All files are available in the central server and the server can communicate with the caching nodes over an error-free broadcast link.
\begin{notation}
We use notation $[n]$ to denote the set of the first $n$ positive integers $\{1,\ldots,n\}$. Similarly, we use $[n]_i$ to denote the set of the first $n$ positive integers larger than $i$, i.e., $\{i+1,i+2,\ldots,i+n\}$.
\end{notation}

The placement of files in the caches can be described as follows. Let $X^n$ be the set of the bits of file $n$. Then, for each $\mathcal{S}\subset[K]$,  set $X^n_\mathcal{S}\subset X^n$ represents the bits of file $n$ that are exclusively cached in caches in $\mathcal{S}$. By definition, subsets $X^n_\mathcal{S}$ are disjoint for different $\mathcal{S}$ and $\bigcup_{\mathcal{S}\subset [K]}X^n_\mathcal{S}=X^n$. Also, define $x^n_{\mathcal{S}}=|X^n_\mathcal{S}|/F$ as the ratio of bits of file $n$ that are  exclusively cached in the caches in $\mathcal{S}$. Then, it follows that $\sum_{\mathcal{S}\subset[K]}x^n_\mathcal{S}=1$  for  every $n\in[N]$. We denote by $\ve{x}$ the vector of all placement parameters $x^n_\mathcal{S}$. 

The server is informed of all  caches' content. For a fixed placement of files in caches, every cache $k\in[K]$ reveals one request for a file $d_k\in[N]$ at each time instant. We refer to $\ve{d}= [d_1,\ldots,d_K]$ as the demand vector which represents the demands of all caches at the current time instant. Similarly, for a subset of caches $\mathcal{S}$, subdemand vector $\ve{d}_\mathcal{S}$ determines the files requested by the caches in $\mathcal{S}$ in the same order as in $\ve{d}$. 

We assume that requests for different files are independent and the request probabilities do not change for different caches. Let $\{p_n\}_{n\in[N]}$ represent the request probabilities of the files. Here, files are sorted in the increasing order of request probabilities, i.e., $n>m$ implies $p_{n}\geq p_{m}$. We refer to the file request probabilities as popularity distribution.  For a demand vector $\ve{d}$ and every $k\in[K]$, the parts of file $d_k$ that are available in cache $k$ are locally delivered to its user. The missing parts are provided by the server over the broadcast channel through a signal of size $R(\ve{d};\mathcal{P},A_D)$ files.   The quantity $R(\ve{d};\mathcal{P},A_D)$ is the delivery rate measured in the equivalent number of files for the demand vector $\ve{d}$, given a specific placement of files $\mathcal{P}$  and a delivery algorithm $A_D$. Placement $\mathcal{P}$ is fixed for all the demand vectors that arrive during the delivery phase. It is required that every cache that has forwarded its request to the server be able to decode the broadcasted signal for the content it requested. We are interested in minimizing $\mathbb{E}_\ve{d}\left(R(\ve{d};\mathcal{P},A_D)\right)$, where the expectation is over the randomness in the demand vector $\ve{d}$.
\vspace*{-.3cm}
\subsection{Delivery Algorithm}
In this work, we apply the \ac{scc} procedure to all user demands for delivery regardless of their popularity levels and the memory allocated to them. The delivery procedure is shown in Algorithm~\ref{alg:del}. 
\begin{algorithm}[t]     
\caption{Delivery by \ac{scc} \cite{Maddah_decentralized:2014}}          
\label{alg:del}          
\begin{algorithmic}[1]                  
\Procedure{Delivery}{$\ve{d};\{X^n\}_{n=1,\ldots,N}$}
\For{$s=1,\ldots,K$}
\For{$\mathcal{S}\subset [K]:|\mathcal{S}|=s$}
\State server sends $\oplus_{k\in \mathcal{S}}X^{d_k}_{\mathcal{S}\backslash k}$
\EndFor
\EndFor
\EndProcedure
\end{algorithmic}
\end{algorithm}
By following Algorithm~\ref{alg:del}, the server transmits messages of the form
\begin{align}\label{eq:serverMess}
\oplus_{k\in\mathcal{S}} X^{d_k}_{\mathcal{S}\backslash k}
\end{align}
for every nonempty $\mathcal{S}\subset[K]$. All the components  $X_{\mathcal{S}\backslash k}$  embedded in the message are zero-padded to the length of the largest component. Hence, the length of the message is
$\max_{k\in\mathcal{S}}|X^{d_k}_{\mathcal{S}\backslash k}|$.\footnote{From a graph theoretic perspective, this message corresponds to XORing the requested subfiles that form a clique in the side information graph \cite[Section~II.A]{Shanmugam:2016} and \cite[Section~I.A]{Bar-Yossef:2011}. Since the set of messages $\oplus_{k\in\mathcal{S}} X^{d_k}_{\mathcal{S}\backslash k}$ delivers all the missing subfiles, it covers all the vertices in the side information graph. Hence, one can see the delivery procedure of \cite{Maddah_decentralized:2014} as a structured way of covering the side information graph vertices with cliques.}  

As mentioned in Section~\ref{sec:related}, Algorithm~\ref{alg:del} contrasts the delivery schemes in \cite{Maddah_nonuniform:2014,Digavi2:2015,Zhang:2015} which are also based on the \ac{scc} procedure but separately apply it to the files with \textit{close} request probabilities. Algorithm~\ref{alg:del} has the advantage that it allows coding among all files regardless of their request probabilities and can result in a smaller delivery rate. To elaborate, message (\ref{eq:serverMess}) delivers every subset of bits in $\{X^{d_k}_{\mathcal{S}\backslash k}\}_{k\in \mathcal{S}}$ to the corresponding requesting cache in $\mathcal{S}$. Given a grouping of files into groups  $l=1,\ldots,L$, if instead of applying the \ac{scc} to the whole demand vector we applied it to subdemand vectors consisting of files in the same popularity group, the exact same subfiles delivered by (\ref{eq:serverMess}) would have been delivered through the set of messages
$\left\{\oplus_{k\in\hat{\mathcal{S}}_l} X^{d_k}_{\mathcal{S}\backslash k}\right\}_{l=1}^L$
where $\hat{\mathcal{S}}_l=\{k\in\mathcal{S}|d_k\in l\text{-th group}\}$. This message has length $\sum_{l=1}^L\max_{k\in\hat{\mathcal{S}}_l}|X^{d_k}_{\mathcal{S}\backslash k}|$ which is lower bounded by $\max_{k\in\mathcal{S}}|X^{d_k}_{\mathcal{S}\backslash k}|$ which is the length of (\ref{eq:serverMess}) with \ac{scc} applied to the whole demand vector.
\vspace*{-.3cm}
\subsection{Formulation of Expected Delivery Rate and Storage}\label{subsec:r_s_form}
To derive optimal placement for \ac{scc}  delivery, we need to formulate the expected delivery rate and the storage used by the placement algorithm in terms of the placement parameters $x^n_\mathcal{S}$.
\subsubsection*{Expected Delivery Rate} 
For Algorithm~\ref{alg:del} as the delivery algorithm,  the delivery load is
\begin{align*}
R(\ve{d};\ve{x})= \sum_{\substack{\mathcal{S}:\mathcal{S}\subset[K] \\ \mathcal{S}\neq\emptyset}}\max_{k\in\mathcal{S}}\,x_{\mathcal{S}\backslash k}^{d_k}
\end{align*}
for a given demand vector $\ve{d}$ and placement parameters $x^n_\mathcal{S}$. 
To formulate the expected delivery rate in terms of the placement parameters, let $R_\mathcal{S}(\ve{d};\ve{x})$ be the  rate required to deliver the subfiles that are exclusively stored in each subset of caches $\mathcal{S}$. Then, the expected rate with respect to randomness in user demands is
\begin{align*}
r(\ve{x})\delequal\mathbb{E}_{\ve{d}}\left(R(\ve{d};\ve{x})\right) = \sum_{\substack{\mathcal{S}:\mathcal{S}\subset\mathcal{K} \\ \mathcal{S}\neq\emptyset}} \mathbb{E}_{\ve{d}}(R_\mathcal{S}(\ve{d};\ve{x})).
\end{align*}

We assumed that the popularity distribution of files is the same for different caches.  We use this symmetry in the demand probabilities of the different caches to simplify the placement formulation by setting $x^n_\mathcal{S}=x^n_s$ for all $\mathcal{S}: |\mathcal{S}|=s$. In other words, for a given file, the portion of bits that is exclusively cached in any subset of caches $\mathcal{S}$ with cardinality $s$ is the same.
Because of the symmetric structure of the placement, $\mathbb{E}_{\ve{d}}(R_\mathcal{S}(\ve{d};\ve{x}))$ is the same for all $\mathcal{S}:|\mathcal{S}|=s$, and it can be denoted by $\bar{R}_s(\ve{x})$. Hence, the average rate can be written as
\begin{align*}
r(\ve{x}) = \sum_{s=1}^K \binom{K}{s}\bar{R}_s(\ve{x}). 
\end{align*}
Let $\mathcal{G}_s$ be the set of all subsets of $[N]$ with cardinality at most $s$.  Let $\pi_s^g$ denote the probability that for a set of caches $\mathcal{S}$ with $|\mathcal{S}|=s$,  $g\in\mathcal{G}_s$ is the set of files in $\ve{d}_\mathcal{S}$. Then, 
\begin{align*}
\bar{R}_s(\ve{x}) = \sum_{g\in\mathcal{G}_s} \pi_s^g \max_{n\in g}x^n_{s-1}
\end{align*}
and therefore, the expected delivery rate is
\begin{align*}
\sum_{s=1}^K \binom{K}{s}\sum_{g\in\mathcal{G}_s} \pi_s^g \max_{n\in g}x^n_{s-1}, 
\end{align*}
which can equivalently be written as
\begin{align*}
\sum_{s=0}^{K-1} \binom{K}{s+1}\sum_{g\in\mathcal{G}_{s+1}} \pi_{s+1}^g \max_{n\in g}x^n_s. 
\end{align*}

\subsubsection*{Storage Used by Placement} 
Under the symmetry conditions derived for placement, the total storage used in each cache is
\begin{align*}
\sum_{n=1}^N\sum_{s=1}^{K}\binom{K-1}{s-1}x_s^n.
\end{align*}
The inner sum  is the storage that is assigned to file $n$ in each cache as for each file $n$, each cache $k$ stores the subfiles $X^n_\mathcal{S}:k\in\mathcal{S}$. There are $\binom{K-1}{s-1}$ subsets of $[K]$ of cardinality $s$ with this property for each file. The outer sum adds up the storage used for all the files in the library.

\subsubsection*{Change of Variable for Placement Parameters}
For simpler exposition of the optimization problems and better interpretability of the results, we find it useful to use the change of variable  
\begin{align}\label{eq:changeXY}
y^n_s=\binom{K}{s}x^n_s.
\end{align} 
Variable $y^n_s$ is the total portion of bits of file $n$ that is cached exclusively in all the $\binom{K}{s}$ different subsets of $[K]$ with cardinality $s$.
As a result, the expected rate and storage can be formulated as functions of the new placement parameters $y^n_s$ as
\begin{align}\label{eq:expectedRateY}
r(\ve{y})&= \sum_{s=0}^{K-1} \frac{K-s}{s+1}\sum_{g\in\mathcal{G}_{s+1}} \pi_{s+1}^g \max_{n\in g}y^n_s,\\
\label{eq:storageUsedY}
m(\ve{y})&=\sum_{n=1}^N\sum_{s=1}^{K}\frac{s}{K}y_s^n.
\end{align}
Notice that the expected rate and the amount of storage used are a convex and a linear function of the placement parameters, respectively.

\section{Formulation of \ac{rmsc} in terms of the placement parameters and characterization of its dual problem}\label{sec:formulation}
\subsection{Formulation of \ac{rmsc}}
Using (\ref{eq:changeXY})-(\ref{eq:storageUsedY}), the problem of finding the storage parameters that minimize the expected delivery rate under cache capacity constraint can be formulated as 
 \begin{mini!}|s|[2]
{\ve{y}}{\sum_{s=0}^{K-1} \frac{K-s}{s+1}\sum_{g\in\mathcal{G}_{s+1}} \pi_{s+1}^g \max_{n\in g}y^n_s\label{eq:objective_rmsc_full}}
{\label{eq:optimProblem_rmsc_full}}
{}
\addConstraint{\sum_{n=1}^N\sum_{s=1}^{K}\frac{s}{K}y_s^n}{\leq M\label{eq:capacityConstraint_rmsc_full}}{}
\addConstraint{\sum_{s=0}^Ky_s^n}{= 1,\;\label{eq:partitionConstraint_rmsc_full}}{n\in[N]}
\addConstraint{y_s^n}{\geq 0,\;\label{eq:signConstraint_rmsc_full}}{n\in[N],\,s=0,\ldots,K,}
\end{mini!}
where constraint (\ref{eq:partitionConstraint_rmsc_full}) follows from $\sum_{\mathcal{S}\subset[K]}x^n_\mathcal{S}=1$ under the same symmetry conditions and change of variable that we used in the derivation of $r(\ve{y})$ and $m(\ve{y})$ in Section~\ref{subsec:r_s_form}. 

\subsection{Duality Framework and Derivation of \acl{jrsm} Problem}\label{sec:pr_du}
Optimization problem (\ref{eq:optimProblem_rmsc_full}) is convex and Slater's condition holds for it\footnote{To check that Slater's condition holds for (\ref{eq:optimProblem_rmsc_full}), notice that there is only one non-affine inequality constraint, (\ref{eq:capacityConstraint_rmsc_full}), in the primal. Consider the point where $y^n_s=1$ for all $n\in[N],s=K$, and $y^n_s=0$ otherwise.  At this point, we have $\sum_{n=1}^N\sum_{s=1}^{K}\frac{s}{K}y_s^n-M=N-M<0$ for any $M<N$, and (\ref{eq:capacityConstraint_rmsc_full}) holds with strict inequality. Also, (\ref{eq:partitionConstraint_rmsc_full}) and (\ref{eq:signConstraint_rmsc_full}) respectively represent equality constraints and affine inequality constraints, and further they are satisfied at the considered point. Hence, we conclude that Slater's condition is satisfied for (\ref{eq:optimProblem_rmsc_full}) with $M<N$ \cite[eq. 5.27]{Boyd:2004}. Notice that if $M\geq N$, the problem is trivial as  all files can be fully cached in every cache of the system. The considered point corresponds to this trivial placement and is optimal.}. Hence, with (\ref{eq:optimProblem_rmsc_full}) as primal, the duality gap between the primal and the corresponding dual problem is zero \cite[Section~5.2]{Boyd:2004}. To derive the dual problem, we form the Lagrangian that accounts for the capacity constraint (\ref{eq:capacityConstraint_rmsc_full}) as
\begin{align*}
L(\ve{y},\gamma)=\sum_{s=0}^{K-1} \frac{K-s}{s+1}\sum_{g\in\mathcal{G}_{s+1}} \pi_{s+1}^g \max_{n\in g}y^n_s+\gamma\left(\sum_{n=1}^N\sum_{s=1}^{K}\frac{s}{K}y_s^n-M\right)
\end{align*}
which results in the Lagrange dual function 
\begin{mini}|s|[2]
{\ve{y}}{L(\ve{y},\gamma)}
{\label{eq:dualFun}}
{g(\gamma)=}
\addConstraint{\sum_{s=0}^Ky_s^n}{= 1,\;}{n\in[N]}
\addConstraint{y_s^n}{\geq 0,\;}{n\in[N],\,s=0,\ldots,K.}
\end{mini}
Then, the corresponding dual problem will be
\begin{maxi*}|s|[2]
{\gamma\geq 0}{g(\gamma).}
{\label{eq:dual_prob}}
{}
\end{maxi*}
By dropping the terms that are independent of the placement parameters, (\ref{eq:dualFun}) has the same minimizers as
 \begin{mini!}|s|[2]
{\ve{y}}{\sum_{s=0}^{K-1} \frac{K-s}{s+1}\sum_{g\in\mathcal{G}_{s+1}} \pi_{s+1}^g \max_{n\in g}y^n_s+\gamma\sum_{n=1}^N\sum_{s=1}^{K}\frac{s}{K}y_s^n\label{eq:objective_jrsm_full}}
{\label{eq:optimProblem_jrsm_full}}
{}
\addConstraint{\sum_{s=0}^Ky_s^n}{= 1,\;\label{eq:partitionConstraint_jrsm_full}}{n\in[N]}
\addConstraint{y_s^n}{\geq 0,\;}{n\in[N],\,s=0,\ldots,K.}
\end{mini!}
We call (\ref{eq:optimProblem_jrsm_full}) the \acrodef{jrsm}[JRSM]{Joint Rate and Storage Minimization} \ac{jrsm} problem, as the objective is to minimize the total bandwidth (expected delivery rate) and storage cost of coded caching. Following the standard interpretation of the Lagrange multipliers, parameter $\gamma$  can be viewed as the relative cost of storage per file. Moreover, since strong duality holds, for each storage capacity $M$, the optimal dual variable $\gamma^*(M)$ determines the pricing of the storage that leads to the same minimizers for the \ac{rmsc} problem (\ref{eq:optimProblem_rmsc_full}) and both the Lagrangian minimization  and \ac{jrsm} problems in (\ref{eq:dualFun}) and (\ref{eq:optimProblem_jrsm_full}). As a result, we derive the optimal solution of \ac{jrsm} in Section~\ref{sec:jrsm} as an intermediate step in solving \ac{rmsc}.
\section{Optimal Solution to \ac{jrsm}}\label{sec:jrsm}
Finding an analytical solution to (\ref{eq:optimProblem_jrsm_full}) is challenging because of the presence of the $\max$ functions that operate over overlapping sets of parameters in the objective. These parameters  are tied together by constraints (\ref{eq:partitionConstraint_jrsm_full}) for different values of $s$. The interplay between the outputs of the $\max$ function applied to the overlapping groups under constraints (\ref{eq:partitionConstraint_jrsm_full}) makes the analysis difficult. 
To facilitate the analysis, we establish a connection between the nonlinear part of (\ref{eq:objective_jrsm_full}) and submodular set functions. This allows us to benefit from the results in submodular function analysis to find structures in the optimal solution to \ac{jrsm}. Appendix~\ref{app:submod} provides a review of submodular functions and the results relevant to our analysis in this paper.

\subsection{An Equivalent Formulation of \ac{jrsm}}
The placement parameters corresponding to $s=0$ are $\{y^n_0\}_{n\in[N]}$, which determine the portion of bits that are not stored in any cache for each file $n$. Also, each set $g\in\mathcal{G}_1$ includes exactly one file, say $g=\{i\}$. Hence, $\max_{n\in g}y^n_0=y^i_0$ and $\pi_0^g=p_i$. Thus, the objective function (\ref{eq:objective_jrsm_full}) can be written as 
\begin{align*}
\sum_{n=1}^NKp_ny^n_0+\sum_{s=1}^{K-1} \left[\frac{K-s}{s+1}\sum_{g\in\mathcal{G}_{s+1}} \pi_{s+1}^g \max_{n\in g}y^n_s+\frac{s}{K}\gamma \sum_{n=1}^Ny_s^n\right]+\gamma\, \sum_{n=1}^Ny_K^n.
\end{align*}
Notice that the first and last sums are in terms of parameters $y^n_0$ and $y^n_K$, respectively, while the summation in the middle accounts for parameters $y^n_s$ for $s\in[K-1]$. 

\begin{lemma}\label{lem:binary}At optimality, $\sum_{n=1}^NKp_ny^n_0+\gamma\sum_{n=1}^Ny_K^n$ can be written as $\sum_{n=1}^N (Kp_n\alpha_n+\gamma(1-\alpha_n))z^n$ where $z^n=y^n_0+y^n_K$, and $\alpha_n=1$ if $Kp_n<\gamma$ and $\alpha_n=0$ if $Kp_n\geq\gamma$.
\end{lemma}
\begin{IEEEproof}For a fixed value of $z^n$, we have $\sum_{n=1}^NKp_ny^n_0+\gamma\sum_{n=1}^Ny_K^n=\gamma\sum_{n=1}^Nz^n+\sum_{n=1}^N(Kp_n-\gamma) y_0^n$. Hence, if $Kp_n<\gamma$, setting $y^n_0=z^n$ and $y^n_K=0$ leads to the smallest objective and if $Kp_n\geq \gamma$, the smallest objective results for $y^n_0=0$ and $y^n_K=z^n$.  
\end{IEEEproof}
\begin{corollary}
For some $m\in\{0,\ldots,N\}$, we have $\alpha_n=1, n\leq m$ and $\alpha_n=0, n>m$.
\end{corollary}

Using Lemma~\ref{lem:binary}, and the fact that
$z^n=1-\sum_{s=1}^{K-1}y_s^n$, we get 
\begin{mini!}|s|[2]
{\tilde{\ve{y}},\ve{\alpha}}{\sum_{s=1}^{K-1}\frac{K-s}{s+1}\!\!\sum_{g\in\mathcal{G}_{s+1}}\!\!\!\pi_s^g\max_{n\in g}y^n_s\!+\!\sum_{n=1}^N\sum_{s=1}^{K-1}\left[\left(\frac{s}{K}-1+\alpha_n\right)\gamma-Kp_n\alpha_n\right]\!y_s^n+l(\ve{\alpha})\label{eq:objective_jrsm_equivalent}}
{\label{eq:optimProblem_jrsm_equivalent}}
{}
\addConstraint{\sum_{s=1}^{K-1}y_s^n}{\leq 1,\;\label{eq:partitionConstraint_jrsm_equivalent}}{n\in[N]}
\addConstraint{y_s^n}{\geq 0,\;}{n\in[N],\,s\in[K-1],}
\addConstraint{\alpha_n}{\in \{0,1\},\;}{n\in[N],}
\end{mini!}
as a problem equivalent to (\ref{eq:optimProblem_jrsm_full}), where $\tilde{\ve{y}}$ is the same as $\ve{y}$, except for parameters $y^n_0$ and $y^n_K$ that are removed, and $l(\ve{\alpha})=K\sum_{n=1}^N\alpha_np_n+\gamma\sum_{n=1}^N(1-\alpha_n)$.

To find structures in the optimal vector $\tilde{\ve{y}}$, assume that the optimal parameters $\alpha_n^*$ are known. Then, the optimization problem for $\tilde{\ve{y}}$ becomes
\begin{mini!}|s|[2]
{\tilde{\ve{y}},t}{t+\sum_{n=1}^N\sum_{s=1}^{K-1}\left[\left(\frac{s}{K}-1+\alpha_n^*\right)\gamma-Kp_n\alpha_n^*\right]\!y_s^n\label{eq:objective_sumodular}}
{\label{eq:optimProblem_submodular}}
{}
\addConstraint{\sum_{s=1}^{K-1}\frac{K\!-\!s}{s\!+\! 1}\!\!\!\sum_{g\in\mathcal{G}_{s+1}}\!\!\!\pi_s^g\max_{n\in g}|y^n_s|\label{eq:ballConstraint_submodular}}{\leq t}
\addConstraint{\sum_{s=1}^{K-1}y_s^n}{\leq 1,\;\label{eq:partitionConstraint_submodular}}{\hspace*{-2cm}n\in[N]}
\addConstraint{y_s^n}{\geq 0,\;\label{eq:boxConstraint_submodular}}{\hspace*{-2cm}n\in[N],\,s\in [K-1].}
\end{mini!}
In constraint (\ref{eq:ballConstraint_submodular}), we used $\max_{n\in g}|y^n_s|$, which is the $l_\infty$-norm instead of $\max_{n\in g}y^n_s$. This does not affect the optimal solution as the two functions are equivalent in the nonnegative orthant specified by (\ref{eq:boxConstraint_submodular}) but makes the LHS in form of the $l_\infty$-norm in Proposition~\ref{pr:mixed_norm} of Appendix~\ref{app:submod}. 

Notice that objective function (\ref{eq:objective_sumodular}) is linear, and  both the objective function and the constraints are in terms of parameters $y^n_s$ for $s\in[K-1],\,n\in[N]$.  For a linear objective function, if the feasible set is convex and bounded with a finite number of extreme points, then there exists an extreme point that is optimal \cite[Section~2.5]{Luenberger:2015}.  In the following, we will show that the feasible set defined by  (\ref{eq:ballConstraint_submodular})-(\ref{eq:boxConstraint_submodular}) satisfies these properties and derive structures in its extreme points.
Any such structure readily implies a structure in at least one optimal solution to (\ref{eq:optimProblem_submodular}).

\subsection{Connection to Submodular Functions}
 To find the extreme points of the region characterized by (\ref{eq:ballConstraint_submodular}), we establish a link to submodular functions. Let us define function
\begin{align*}
f_{\text{c}}(\tilde{\ve{y}})\delequal\sum_{s=1}^{K-1}\frac{K\!-\!s}{s\!+\! 1}\!\!\!\sum_{g\in\mathcal{G}_{s+1}}\!\!\!\pi_s^g\max_{n\in g}|y^n_s|.
\end{align*}
The subscript c is used to highlight that this function returns the average rate due to the delivery of the bits that are cached in at least one of the caches in the system.
We show that $f_{\text{c}}(\tilde{\ve{y}})$ is the Lov\'{a}sz extension of a submodular set function. For that, consider the set $[(K-1)N]$. For each $s\in[K-1]$, objects ${(s-1)N+1},\ldots,{sN}$ correspond to files $1,\ldots,N$, respectively. Notice that these objects belong to $[N]_{(s-1)N}$. 

To define the corresponding set function, let us introduce the following for any $s\in[K-1]$ and $g\in\mathcal{G}_{s+1}$:
\begin{itemize}
\item Operator $u(s,g)$ that gives the set $\tilde{g}=\{{
(s-1)N+n}\mid n\in g\}$ as output. This is basically a mapping from the files in $g$ and set sizes $s$ to the objects in $[(K-1)N]$. Notice that any resulting set $\tilde{g}$ is a subset of $[N]_{(s-1)N}$ for exactly one $s$.
\item Sets $\tilde{\mathcal{G}}_{s+1}=\{u(s,g)\mid g\in\mathcal{G}_{s+1}\}$ and $\tilde{\mathcal{G}}=\bigcup_{s\in[K-1]}\tilde{\mathcal{G}}_{s+1}$.
\item The inverse operators $s^{-1}(\tilde{g})$ and $g^{-1}(\tilde{g})$ that for $\tilde{g}\in\tilde{\mathcal{G}}$ return the unique $s$ that satisfies $\tilde{g}\subset[N]_{(s-1)N}$, and the set $g=\{n\mid {s^{-1}(\tilde{g})N+n}\in \tilde{g}\}$, respectively.
\item Weights 
\begin{align}\label{eq:lovaz_f}
\eta_{\tilde{g}}=\frac{K-s^{-1}(\tilde{g})}{s^{-1}(\tilde{g})+1}\pi_{s^{-1}(\tilde{g})}^{g^{-1}(\tilde{g})}
\end{align}for all $\tilde{g}\in\tilde{\mathcal{G}}$. Notice that when $|\tilde{g}|=1$, $g^{-1}(\tilde{g})=\{i\}$ which is a singleton. In that case, $\pi_{s^{-1}(\tilde{g})}^{g^{-1}(\tilde{g})}=p_i$.
\end{itemize}
Using the operators and parameters defined above, $f_{\text{c}}(\tilde{\ve{y}})$ can be written as
\begin{align}\label{eq:lovaz_f}
f_{\text{c}}(\tilde{\ve{y}})=\sum_{\tilde{g}\in\tilde{\mathcal{G}}}\eta_{\tilde{g}}\max_{n\in g^{-1}(\tilde{g})}|y^n_{s^{-1}(\tilde{g})}|.
\end{align}
Notice that (\ref{eq:lovaz_f}) has the form of the norm function in Proposition~\ref{pr:mixed_norm} and as a direct consequence we have the following proposition:
\begin{proposition}\label{lem:subObj} Function $f_{\text{c}}(\tilde{\ve{y}})$ is a norm   and is the Lov\'{a}sz extension of the submodular function 
\begin{align}
F_{\text{c}}(A)=\sum_{\tilde{g}\in\tilde{\mathcal{G}}:A\cap \tilde{g}\neq\emptyset}\eta_{\tilde{g}},
\end{align} 
where $A\subset [(K-1)N]$.
\end{proposition}
From Proposition~\ref{lem:subObj}, one concludes that constraint (\ref{eq:ballConstraint_submodular}) characterizes a norm-ball of radius $t$. 

For $A\subset[N]_{(s-1)N}$, let us define $P(A)=\sum_{n\in g^{-1}(A)}p_n$. Then, for the extreme points of the norm-ball, we have the following lemma.
\begin{lemma}\label{lem:extremePoints} The extreme points of the  norm-ball $f_{\text{c}}\leq t$ are of the form 
\begin{align}\label{eq:extremePoints}
\frac{t}{\frac{K-s}{s+1}\left[1-(1-P(A))^{s+1}\right]}\ve{v},
\end{align} 
where vector $\ve{v}\in \{-1, 0, 1\}^{KN}$, $\text{Supp}(\ve{v}) = A$, and  set $A$ is a subset of $[N]_{(s-1)N}$ for an $s\in[K-1]$.
\end{lemma}
\begin{IEEEproof}
Based on Proposition~\ref{eq:ext_ball}, the extreme points of the unit ball $f_{\text{c}}\leq 1$ are closely connected to the set of stable inseparable subsets of $[(K-1)N]$ with regard to $F_{\text{c}}$. We first argue that all subsets of $[(K-1)N]$ are stable. Consider a set $A\subset[(K-1)N]$. Augment $A$ with a new object $i$ to get $A\cup\{i\}$. Without loss of generality, let $s^
{-1}(\{i\})=\hat{s}$. Since $\tilde{g}=\{i\}$ belongs to $\mathcal{G}_{\hat{s}+1}$ with $\eta_{\tilde{g}}>0$ and it does not intersect with $A$, we have $f_{\text{c}}(A\cup\{i\})>F_{\text{c}}(A)$. Hence, any set $A\subset[(K-1)N]$ is stable with respect to $F_{\text{c}}$. Consequently,  every subset of $[N]_{(s-1)N}$ for $s\in[K-1]$ is also stable.

To find the inseparable sets, consider $A\subset[(K-1)N]$. Let $B_{s}=\{i\in A \mid s^{-1}(\{i\})=s\}$. A necessary condition for $A$ to be inseparable is to have only one nonempty $B_{s}$. To show this, partition $A$ to subsets $B_s,s\in[K-1]$. Notice that each group $\tilde{g}\subset\tilde{\mathcal{G}}$ is a subset of exactly one $[N]_{(s-1)N},s\in[K-1]$.  Hence, if two or more subsets $B_s$ are nonempty, then $F_{\text{c}}(A)=\sum_{B_s\neq \emptyset}f_{\text{c}}(B_s)$ and $A$ is separable. 
Now, consider the case where only one $B_s$, say $B_{\hat{s}}$, is nonempty.  In this case, $A\subset [N]_{(\hat{s}-1)N}$ and $A$ can only have nonempty intersections with sets $\tilde{g}\in\tilde{\mathcal{G}}_{\hat{s}+1}$. 
Since $\hat{s}\geq 1$, for any partitioning of $A$ to $P_1,\ldots,P_J$ for some $J$, there is at least one group $\tilde{g}\in\tilde{\mathcal{G}}_{\hat{s}+1}$ with $|\tilde{g}|\geq 2$ that intersects with both $P_i$ and $P_j$ for every pair $i\neq j$. Hence, $F_{\text{c}}(A)<\sum_{i=1}^Jf_{\text{c}}(P_i)$. 
As a result, the set of all stable inseparable subsets of $[(K-1)N]$ with regard to $F_{\text{c}}$ is $\mathcal{A}=\{A\mid A\subset [N]_{(s-1)N},s\in[K-1]\}$.

According to Proposition~\ref{eq:ext_ball},  the support of every extreme point of the norm-ball of $f_{\text{c}}$ belongs to $\mathcal{A}$. Further, the nonzero entries of the extreme point vector that corresponds to $A\in\mathcal{A}$ is either of $\pm 1/F_{\text{c}}(A)$. Using Proposition~\ref{lem:subObj}:
\begin{align}\label{eq:f_c}
F_{\text{c}}(A)&=\sum_{\tilde{g}\subset\tilde{\mathcal{G}}:A\cap \tilde{g}\neq\emptyset}\eta_{\tilde{g}}
=\sum_{\substack{\tilde{g}\subset\tilde{\mathcal{G}}_{s^{-1}(A)}: \\A\cap \tilde{g}\neq\emptyset}}\eta_{\tilde{g}}
=\frac{K-s^{-1}(A)}{s^{-1}(A)+1}\left(1-\left(1-P(A)\right)^{s^{-1}(A)+1}\right)
\end{align}
where we used the facts that 1) for $A\in\mathcal{A}$, all entries of $A$ belong to only one $[N]_{(s-1)N}$, so $s^{-1}(A)$ and $g^{-1}(A)$ are well defined, 2) $\eta_{\tilde{g}}=\frac{K-s^{-1}(\tilde{g})}{s^{-1}(\tilde{g})+1}\pi_{s^{-1}(\tilde{g})}^{g^{-1}(\tilde{g})}$ and 3) $ \sum_{\substack{\tilde{g}\subset\tilde{\mathcal{G}}_{s^{-1}(A)}: \\A\cap \tilde{g}\neq\emptyset}}\eta_{\tilde{g}}$ equals  the probability of having a demand vector with at most $s^{-1}(A)+1$ distinct files from $[N]$ that has at least one file in $g^{-1}(A)$. The use of $\mathcal{A}$ and (\ref{eq:f_c}) in  Proposition~\ref{eq:ext_ball} and scaling the radius of the ball from 1 to $t$ results in (\ref{eq:extremePoints}).
\end{IEEEproof}

\begin{corollary}
For an extreme point $\tilde{\ve{y}}$ of the norm-ball defined by (\ref{eq:ballConstraint_submodular}), for all $y^n_s>0$, we have $s=\hat{s}$, for exactly one $\hat{s}\in[K-1]$.
\end{corollary}

\begin{theorem}\label{th:optimal} There is an optimal solution to the \ac{jrsm} problem in (\ref{eq:optimProblem_jrsm_full}) which is of form
\begin{align}\label{eq:optimal_jrsm}
(y^n_s)^*= \begin{cases}
               1,             & s=0,\,n\in[n^*];\,s=s^*,n\in[N-n^*]_{n^*},            \\
               0,             & \text{otherwise},
           \end{cases}
\end{align}
for some $s^*\in[K]$ and some $n^*\in\{0,\ldots,N\}$.\footnote{Notice that $n^*=0$ corresponds to the case where $y^n_{s^*}=1$ for all $n\in[N]$.}
\end{theorem}
\begin{IEEEproof}
See Appendix~\ref{app:th1_proof}.
\end{IEEEproof}

Theorem~\ref{th:optimal} implies that for every $\gamma\geq 0$ there exists an optimal solution to the \ac{jrsm} problem that is integral. For better illustration, we write such an optimal vector $\ve{y}$ in the matrix form $Y$ as follows. Matrix $Y$ has $N$ rows corresponding to the files and $K+1$ columns corresponding to the cardinality of subsets of caches. In particular, $Y_{n,s}=y^n_s,n\in[N], s=0,\ldots,K$. Based on the structures found for $y^n_s$ in Theorem \ref{th:optimal}, $Y$ is of form
{
\begin{align}\label{eq:matrix}
Y=\kbordermatrix{
  &0& 1 & \dots  & s^*-1  & {s}^* & s^*+1  & \dots  & K \\
1 &\;\;\textbf{1}\;\; &\;\;0\;\; &\;\;\dots\;\; & \;\;0\;\; & \;\;\textbf{0}\;\; & \;\;0\;\; & \;\;\dots\;\;  & \;\;0\;\; \\
2 &\;\;\textbf{1}\;\; &\;\;0\;\; &\;\;\dots\;\; & \;\;0\;\; & \;\;\textbf{0}\;\; & \;\;0\;\; & \;\;\dots\;\;  & \;\;0\;\; \\
\vdots &\;\;\textbf{\vdots}\;\; &\;\;\vdots\;\; &\;\;\ddots\;\; & \;\;\textbf{\vdots} \;\;& \;\;\vdots\;\; & \;\;\vdots\;\; & \;\;\ddots\;\; & \;\;\vdots \;\;\\
n^*-1 &\;\;\textbf{1}\;\; &\;\;0\;\; &\;\;\dots\;\; & \;\;0\;\; & \;\;\textbf{0}\;\; & \;\;0\;\; & \;\;\dots\;\;  & \;\;0\;\; \\
{n}^*&\;\;\textbf{0}\;\; &\;\;0\;\; &\;\;\dots\;\; & \;\;0\;\; & \;\;\textbf{1}\;\; & \;\;0\;\; & \;\;\dots\;\;  & \;\;0\;\; \\
n^*+1 &\;\;\textbf{0}\;\; &\;\;0\;\; &\;\;\dots\;\; & \;\;0\;\; & \;\;\textbf{1}\;\; & \;\;0\;\; & \;\;\dots\;\;  & \;\;0\;\; \\
\vdots &\;\;\textbf{\vdots}\;\; &\;\;\vdots\;\; &\;\;\ddots\;\; & \;\;\vdots \;\;& \;\;\vdots\;\; & \;\;\textbf{\vdots}\;\; & \;\;\ddots\;\; & \;\;\vdots \;\;\\
N &\;\;\textbf{0}\;\; &\;\;0\;\; &\;\;\dots\;\; & \;\;0\;\; & \;\;\textbf{1}\;\; & \;\;0\;\; & \;\;\dots\;\;  & \;\;0\;\; },
\end{align}}
i.e., i) all entries are either 0 or 1, ii)  each row has exactly one entry 1, iii) at most one column with index $s\geq 1$ has nonzero entries, iv) for that column all entries are 0 for rows $1,\ldots,n^*-1$ and 1 for rows $n^*,\ldots,N$, for some $n^*\in[N]$. As a result, for all values of $\gamma\geq 0$ there is an optimal solution with matrix form (\ref{eq:matrix}). Hence, we have the following corollary:
\begin{corollary}\label{cor:finite}
There exists a finite set $\mathcal{Y}^*$ of vectors that correspond to matrices of form (\ref{eq:matrix}) where $|\mathcal{Y}^*|\leq KN+1$ and that set includes at least one optimal solution to the \ac{jrsm} problem (\ref{eq:optimProblem_jrsm_full}) for every $\gamma\geq 0$.\footnote{We show in Appendix~\ref{app:th2_proof} that for specific values of $\gamma$, there are infinite number of solutions to the \ac{jrsm} problem.}
\end{corollary} 

\section{Optimal Solution to \ac{rmsc}}\label{sec:analysis_rmsc}
\subsection{Optimal Solution of \ac{rmsc} in Terms of Optimal \ac{jrsm} Solution}
Assuming that the optimal dual parameter $\gamma^*$ is known, we derived structures in the minimizers of the Lagrangian or equivalently in the optimal solution of \ac{jrsm}. The derived structures limited the search space for the optimal \ac{jrsm} solution to $KN+1$ vectors specified by Theorem~\ref{th:optimal}. In this section, we derive the optimal solution to \ac{rmsc} by building on the results we derived for the solution of \ac{jrsm} in Theorem~\ref{th:optimal}. For that, let us define two sets as follows:
\begin{definition}\label{def:m_r}
Sets $\mathcal{M}$ and $\mathcal{R}$ are  finite sets  defined by storage values $\{m(\ve{y})\mid \ve{y}\in\mathcal{Y}^*\}$ and expected rates $\{r(\ve{y})\mid \ve{y}\in\mathcal{Y}^*\}$, respectively. 
\end{definition}
To characterize the solution of \ac{rmsc}, we take the following two steps. First, we assume that set $\mathcal{Y}^*$ and consequently set $\mathcal{M}$ are  known. Based on this assumption, we derive the optimal dual parameter $\gamma^*$ as a function of storage capacity $M$ in the primal problem. Second, we use the derived $\gamma^*$-$M$ relationship to find set $\mathcal{Y}^*$ and derive the optimal solution to \ac{rmsc}.

\begin{figure}
\centering
\includegraphics[width=0.65\linewidth]{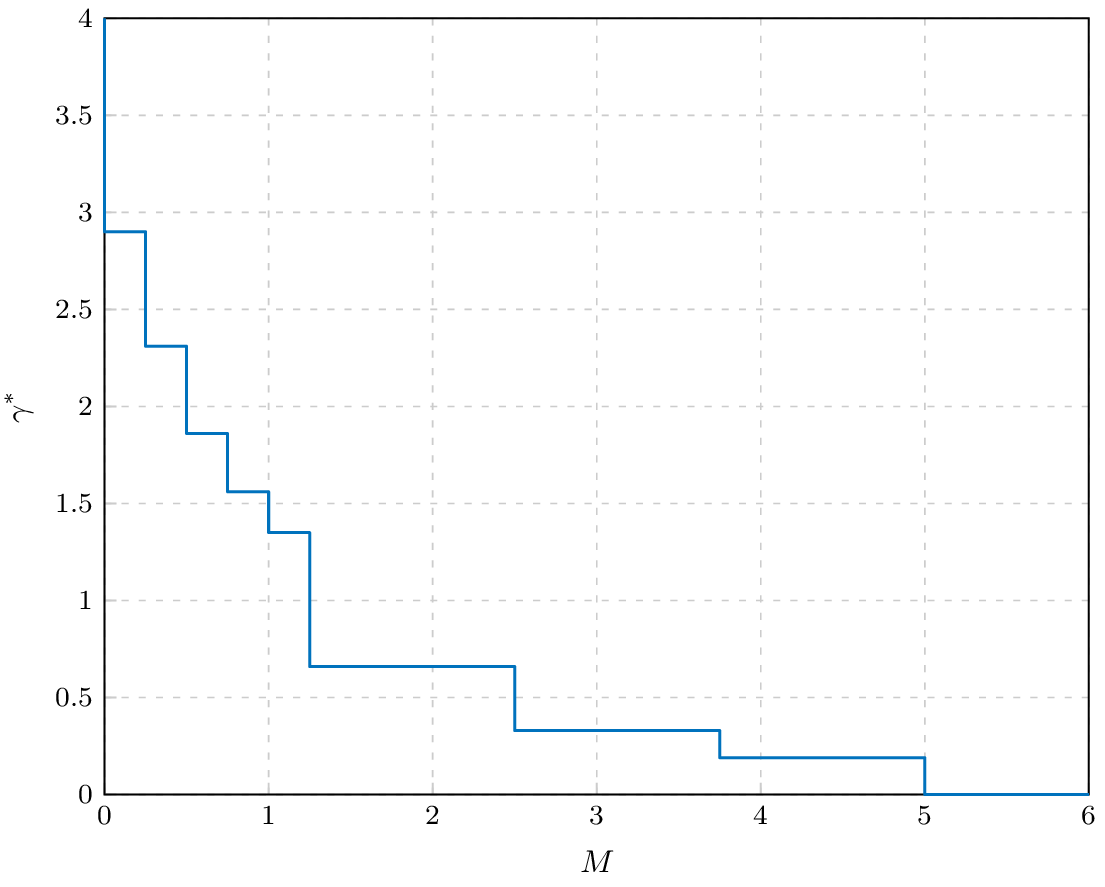}
\caption{Here, $N=5$, $K=4$, and the popularity distribution is Zipf with parameter $\alpha=1$.}\label{fig:Msample}
\end{figure} 
\begin{lemma}\label{lem:gamma_M} The optimal dual parameter $\gamma^*$ and the storage capacity $M$ in the primal \ac{rmsc} problem satisfy the following:
\begin{enumerate}
\item Parameter $\gamma^*$ is non-increasing in $M$;
\item For certain storage capacities $M$, a continuum of dual parameters $\gamma^*$ are optimal;
\item For every two consecutive values $M_1,M_2\in\mathcal{M}, M_1<M_2$,  any $M\in[M_1,M_2]$ leads to the same dual optimal parameter $\gamma^*$. 
\end{enumerate}
\end{lemma}
\begin{IEEEproof}
See Appendix~\ref{app:lem_gamma_M}.
\end{IEEEproof}
Lemma~\ref{lem:gamma_M} implies a stairwaise relationship between the optimal dual parameter $\gamma^*$ and the storage capacity $M$ in the primal problem.  
An illustration of this relationship is shown in Fig.~\ref{fig:Msample}.  The fact that $\gamma^*$ is non-increasing in $M$ is in agreement with the interpretation of the Lagrange multiplier $\gamma^*$ as the (relative) price per unit of storage \cite{Boyd:2004}: as more storage becomes available, the storage price remains the same or decreases.
The second point in the lemma corresponds to the vertical line segments in Fig.~\ref{fig:Msample}. Based on Definition~\ref{def:m_r},  set $\mathcal{M}$ which is derived from $\mathcal{Y}^*$ is finite and has at most $KN+1$ members. However, $\gamma$ is a continuous variable and for every $\gamma\geq 0$ there is an optimal solution to \ac{jrsm} in $\mathcal{Y}^*$. Hence, an interval of $\gamma$ values must map to the same $M$. 
Third,  a range of values of $M$ are mapped into the same $\gamma^*$. Notice that parameter $M$ in the primal problem can take any nonnegative value, while $\mathcal{M}$ is a set of discrete values and is of finite size. Since every $M\geq 0$ corresponds to at least one optimal dual parameter $\gamma$, then a continuum of values for $M$ must map to the same $\gamma^*$.
We show that parameter $\gamma^*$ and the two solutions $\ve{y}_1,\ve{y}_2\in\mathcal{Y}^*$ that lead to the two endpoints $(m(\ve{y}_1),\gamma^*)$ and $(m(\ve{y}_2),\gamma^*)$ of the line segment are related by $\gamma^*=\frac{r(\ve{y}_1)-r(\ve{y}_2)}{m(\ve{y}_2)-m(\ve{y}_1)}$. Notice that $m(\ve{y}_1),m(\ve{y}_2)\in\mathcal{M}$ and  $r(\ve{y}_1),r(\ve{y}_2)\in\mathcal{R}$. In particular, if we sort members of $\mathcal{M}$ in increasing order as $0=M_0< M_1 <M_2<\ldots <M_{l}=N$, then rates $R_i$ that correspond to storage values $M_i$ follow ordering $K=R_0>R_1>\ldots>R_l=0$. Hence
\begin{align*}
\gamma^*(M)=
\left\{
	\begin{array}{lll}
	[\frac{R_{i}-R_{i+1}}{M_{i+1}-M_{i}},\frac{R_{i-1}-R_{i}}{M_{i}-M_{i-1}}],  &  M = M_i \\
		\frac{R_{i-1}-R_{i}}{M_{i}-M_{i-1}},  &  M_{i-1}< M <M_{i}
	\end{array}
\right.
\end{align*}
with $\gamma^*(M_0=0)=[\frac{K-R_1}{M_1},+\infty]$ and $\gamma^*(M_l=N)=[0,\frac{R_{l-1}}{N-M_{l-1}}]$.

The next theorem determines the relationship between the optimal solution of  \ac{rmsc} and the optimal solution of \ac{jrsm} which was derived in Theorem~\ref{th:optimal}.

\begin{theorem}\label{th:optimal_rmsc}
The $\ac{rmsc}$ problem (\ref{eq:optimProblem_rmsc_full}) has an optimal solution
\begin{align}\label{eq:optimal_rmsc}
y^*_{\text{\ac{rmsc}}}(M)=
\left\{
	\begin{array}{ll}
		y^*_{\text{\ac{jrsm}}}(M),  &  M \in \mathcal{M} \\
		\frac{M_\text{u}-M}{M_\text{u}-M_\text{l}} y^*_{\text{\ac{jrsm}}}(M_\text{l})+ \frac{M-M_\text{l}}{M_\text{u}-M_\text{l}}y^*_{\text{\ac{jrsm}}}(M_\text{u}), & M\not\in\mathcal{M}
	\end{array}
\right.
\end{align}
where $y^*_{\text{\ac{jrsm}}}(m)$ is the optimal solution of \ac{jrsm} of the form in Theorem~\ref{th:optimal} that uses storage $m$, and $M_\text{l}$ and $M_\text{u}$ are the largest element smaller than $M$ and smallest element larger than $M$ in $\mathcal{M}$, respectively.
\end{theorem}
\begin{IEEEproof}
See Appendix~\ref{app:th2_proof}
\end{IEEEproof}
\paragraph*{Optimality of Memory Sharing} Theorem~\ref{th:optimal_rmsc} essentially extends a result known for the optimal solution of \ac{rmsc} for uniform demands to the general case where demands can be nonuniform. To elaborate, it has been shown that for uniform demands, if $M\in\{\frac{N}{K},2\frac{N}{K},\ldots,K\frac{N}{K}\}$, then the optimal solution of \ac{rmsc} is in the form in (\ref{eq:matrix}) for some $s^*=0,\ldots,K$ and $n^*=1$ \cite{ours:arxiv,Caire:2017,Yu:2017}. In particular, for uniform demands $\mathcal{M}_{\text{uniform}}=\{\frac{N}{K},2\frac{N}{K},\ldots,K\frac{N}{K}\}$. For other values of $M$, the optimal solution could be obtained by memory sharing between the two values of storage in $\mathcal{M}_{\text{uniform}}$ closest to $M$. Theorem~\ref{th:optimal_rmsc} shows that the same result is valid for nonuniform demands except for the fact that $n^*$ might be any value between $1$ and $N$, depending on the popularity distribution of files. As a result, we propose the following terminology:
\begin{definition}
For a given number of caches and popularity distribution of files, we call set $\mathcal{M}$ the set of base-cases of the \ac{rmsc} problem. 
\end{definition}
Based on Theorem~\ref{th:optimal}, for base-cases of \ac{rmsc}, there exists an optimal solution which is integral. Also, from Theorem~\ref{th:optimal_rmsc},  for other storage capacities, the optimal  solution to \ac{rmsc} can be obtained by memory sharing between the solutions of two certain base-cases.
\subsection{Algorithm to Derive $\mathcal{M}$}
We derive an algorithm with a worst-case complexity of $O(K^2N^2)$ to find set $\mathcal{Y}^*$ and consequently $\mathcal{M}$ for any given number of caches and popularity distribution of files. With $\mathcal{M}$ determined, Theorem~\ref{th:optimal_rmsc} analytically solves the \ac{rmsc} problem for any cache capacity.
\begin{algorithm}[t]     
\caption{Procedure to Determine the Set of Base-Cases $\mathcal{M}$}     
\label{alg:optimize}          
\begin{algorithmic}[1]                  
\Procedure{base}{$K,N,\{p_n\}$}
\State \# Calculate Storage and Rate for the $KN+1$ matrices of form (\ref{eq:matrix}) 
\State $Y_o\leftarrow\ve{0}_{N\times (K+1)}$, $M_0\leftarrow 0$, $R_0\leftarrow K$
\For{$s=1,\ldots,K$}
\For{$n=1:N$}
\State $i\leftarrow (s-1)N+n$
\State $Y_i\leftarrow\ve{0}_{N\times (K+1)}$,  $Y_i(1:n-1,0)\leftarrow 1$, $Y_i(n:N,s)\leftarrow 1$
\State $M_i\leftarrow m(Y)$
\State $R_i\leftarrow r(Y)$
\EndFor 
\EndFor
\State $(M,R,Y)\leftarrow \text{sort}_M(M,R,Y)$ \# relabel $(M_i,R_i,Y_i) $ tuples in increasing order of $M_i$
\State \# Build $\mathcal{M}$, $\mathcal{R}$ and $\mathcal{Y}$ by keeping solutions that outperform memory sharing between other cases
\State $(\mathcal{Y}_0,\mathcal{M}_0,\mathcal{R}_0)\leftarrow(\ve{0}_{N\times (K+1)},0,K)$
\State $c\leftarrow 0$
\For{$i=1,\ldots,NK+1$}
\For{$j=i+1:NK+1$}
\State $R_{\text{msh}}\leftarrow\frac{M_j-M_i}{M_j-\mathcal{M}_c}\mathcal{R}_c + \frac{M_i-M_j}{M_j-\mathcal{M}_c}R_j$
\If{$R_i<R_{\text{msh}}$} 
\State $c\leftarrow c+1$
\State $(\mathcal{Y}_c,\mathcal{M}_c,\mathcal{R}_c)\leftarrow (Y_i,M_i,R_i)$
\State \textbf{break}
\EndIf
\EndFor 
\EndFor
\EndProcedure
\end{algorithmic}
\end{algorithm}

To find $\mathcal{Y}^*$, we need to search over the $KN+1$ possibilities for $\ve{y}^*$ of form (\ref{eq:optimal_jrsm}). For each such vector $\ve{y}$, the storage it uses  can be written as a convex combination of the storage used by two other vectors $\ve{y}_1$ and $\ve{y}_2$ that satisfy $m(\ve{y}_1)\leq m(\ve{y})$ and $m(\ve{y}_2)\geq  m(\ve{y})$. In other words, $m(\ve{y})=m(\theta\ve{y}_1+(1-\theta)\ve{y}_2), 0\leq \theta\leq 1$.\footnote{except for the two vectors with $m(\ve{y})\in\{0,N\}$.} If for such $\ve{y}_1,\ve{y}_2$, we further have $r(\ve{y})\geq\tilde{m}(\ve{y})$, then $\ve{y}$ does not belong to $\mathcal{Y}^*$. Hence, by removing such vectors form the $KN+1$ possibilities, the remaining set of vectors constitutes $\mathcal{Y}^*$. 
The \textsc{base} procedure in Algorithm~\ref{alg:optimize} implements this process by starting from $Y$ with all entries in  the first column equal to 1. This correspond to storage $0$ and rate $K$ and belongs to $\mathcal{Y}^*$. It then proceeds to the next $\ve{y}$ with the smallest storage value. It checks whether it outperforms memory sharing between the $\ve{y}$ that is already in $\mathcal{Y}^*$  with the largest storage and every remaining vector that uses more storage compared to $\ve{y}$. If that is the case, it adds the new vector to $\mathcal{Y}^*$, otherwise drops the vector and proceeds to the next vector.

\begin{figure}
\centering
\includegraphics[width = 0.8\textwidth]{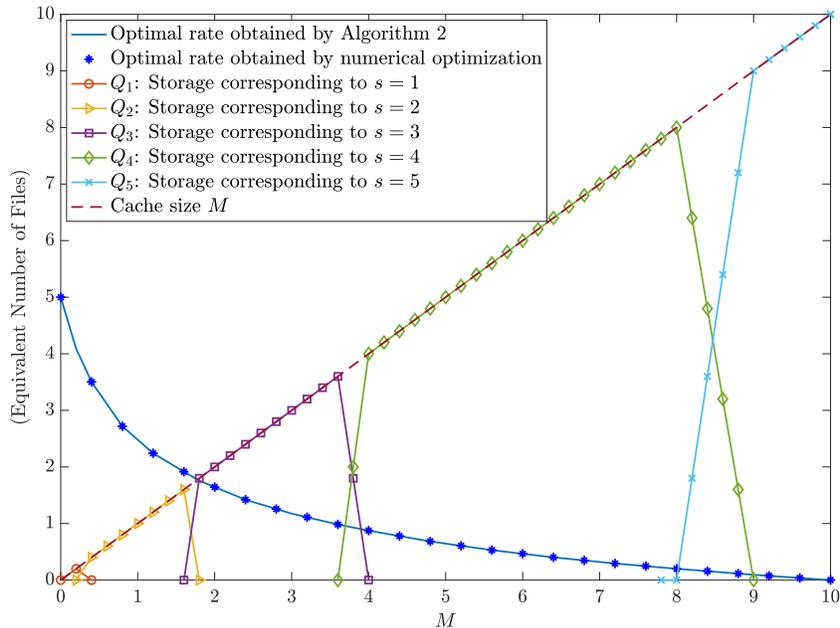}
\caption{The effect of cache size on expected delivery rate and the amount of storage used to cache subsets of files that are exclusively stored in subsets of caches with cardinalities $1,\ldots,K$ for $K=5$, $N=10$. Here, the popularity of files follows a Zipf distribution with parameter $1.4$.}\label{fig:result}
\end{figure}

Fig.~\ref{fig:result} shows the expected delivery rate of the proposed method versus the cache capacity for a nonuniform distribution of files that follows a Zipf density with parameter 1.4. The expected rate is once calculated based on the solution obtained by Algorithm~\ref{alg:optimize} and once by  solving \ac{rmsc} numerically. We observe that the resulting optimal rates are in complete agreement. Fig.~\ref{fig:result} also shows the amount of storage used to cache subsets of files that are exclusively stored in subsets of caches with different cardinalities $s\in[K]$. In other words, for each $s$, it shows $Q_s\delequal\sum_{n=1}^N\frac{s}{K}y^n_s$ as a function of the cache capacity. As we expect from our analysis, either one or two values of $Q_s$ can be positive for each choice of $M$. 
\section{Conclusion}\label{sec:con}
 In this paper, we applied the structured clique cover delivery algorithm that was proposed for decentralized coded caching to deliver files with nonuniform request probabilities. We fully characterized the structure of the optimal placement parameters. We showed that for a finite set of cache capacities, called base-cases, the optimal placement follows the two group strategy that does not cache a subset of less popular files, but treats the other files  selected for caching identically regardless of their popularities. A polynomial time procedure was also proposed to derive this set of cache capacities. We further showed that the optimal placement parameters for other storage capacities can be obtained by memory sharing between certain base cases. In this scenario, a grouping of files into two or three groups is optimal.

\appendices
\counterwithin{proposition}{section}
\counterwithin{lemma}{section}
\section{Review of Submodular Functions and Analysis}\label{app:submod}
\subsection{Review of Submodular Functions and Relevant Results}
We review the definition of a submodular set function and present the results that are related to our analysis in Section~\ref{sec:jrsm}. An extended discussion can be found in \cite{bach:2010}.
\begin{definition}
Let $V = \{1,\ldots, p\}$ be a set of $p$ objects. For $\ve{w}\in \mathbb{R}^p$, $\text{Supp}(\ve{w}) \subset V$ denotes the support of $\ve{w}$, defined as
$\text{Supp}(\ve{w}) = \{j \in V, w_j \neq 0\}$.
\end{definition}
\begin{definition}
\textbf{(Submodular function)} A set-function $F : 2^V \rightarrow
\mathbb{R}$ is submodular if and only if, for all subsets $A, B \subset V$, we have
$F(A) + F(B) \geq F(A \cup B) + F(A \cap B)$.
\end{definition} 

\begin{definition} \textbf{(Lov\'{a}sz extension)} Given a set-function $F$ such that $F(\emptyset) = 0$, the Lov\'{a}sz extension $f : \mathbb{R}^p_+ \rightarrow \mathbb{R}$ of $F$ is defined as 
\begin{align*}
f(\ve{w}) = \sum_{k=1}^pw_{j_k} F(\{j_1, \ldots , j_k\}) - F(\{j_1, \ldots , j_{k-1}\}),
\end{align*}
where $\ve{w} \in \mathbb{R}^p_+$, ($j_1, \ldots , j_p)$ is a permutation such that   $w_{j_1} \geq \ldots \geq w_{j_p}$.
\end{definition} 

Consider vector $\ve{\delta}\in\{0, 1\}^p$ as the indicator vector for subset $A\subset V$, i.e, for $i\in V$, $\delta_i=1$ if and only if $i\in A$. Consequently, $A$ is the support of $\ve{\delta}$. Notice that for the Lov\'{a}sz extension, $f(\ve{\delta}) = F(\text{Supp}(\ve{\delta}))$. Hence, $f$ can be seen as an extension of $F$ from vectors in $\{0, 1\}^p$ to all vectors in $\mathbb{R}^p_+$. The Lov\'{a}sz extension $f$ has the following properties: 1)  it is piecewise-linear, 2) when $F$ is submodular, $f$ is convex, and 3) minimizing $F$ over subsets, i.e., minimizing $f$ over $\{0, 1\}^p$,
is equivalent to minimizing $f$ over $[0, 1]^p$.

\begin{definition} \textbf{(Stable Sets)} A set $A$ is stable if it cannot be augmented without increasing $F$, i.e., if for all
sets $B \supset A$, $B \neq A$, then $F(B) > F(A)$. 
\end{definition}

\begin{definition} \textbf{(Separable Sets)} A set $A$ is separable if we can find a partition of $A$ into $A = B_1\cup\ldots\cup B_k$ such that
$F(A) = F(B_1) + \ldots + F(B_k)$. A set $A$ is inseparable if it is not separable.
\end{definition}

\begin{proposition}\label{pr:mixed_norm}\!\!\cite[Section~4.1]{bach:2010} Let $G$ be a (overlapping) grouping of the objects in $V$ and $d$ be a nonnegative set-function. For  $\ve{w}\in \mathbb{R}^p$, function $\Omega(\ve{w}) = \sum_{G\subset V}d(G)\norm{\ve{w}}_\infty$ is a norm if $\cup _{G,d(G)>0}G = V$ and it corresponds to the nondecreasing submodular function $F(A)=\sum_{\substack{G:A\cap G\neq\emptyset}}d(G)$.
\end{proposition}

\begin{proposition}\label{eq:ext_ball} \cite[Proposition~2]{bach:2010}) The extreme points of the unit ball of $\Omega$ are the vectors $\frac{1}{F(A)}\ve{v}$, with $\ve{v} \in \{-1, 0, 1\}^p$, $\text{Supp}(\ve{v}) = A$ and $A$ a stable inseparable set.
\end{proposition}

\section{Proof of Theorem~\ref{th:optimal}}\label{app:th1_proof}
To prove Theorem~\ref{th:optimal}, we first prove the following lemma.
\begin{lemma}\label{lem:extPointsFeas} The extreme points of the region $[f_{\text{c}}\leq t]^+$ defined by  (\ref{eq:ballConstraint_submodular}) and (\ref{eq:boxConstraint_submodular}) are the origin and points of the form 
\begin{align}\label{eq:extPointsFeas}
\frac{t}{\frac{K-s}{s+1}\left[1-(1-P(A))^{s+1}\right]}\ve{v},
\end{align} 
where vector $\ve{v}\in \{0, 1\}^{KN}$, $\text{Supp}(\ve{v}) = A$, and set $A$ is a subset of $[N]_{(s-1)N}$ for an $s\in[K-1]$.
\end{lemma}
\begin{IEEEproof}
To obtain the extreme points of $[f_{\text{c}}\leq t]^+$ we begin with the extreme points of the norm-ball $f_{\text{c}}\leq t$ and remove the ones that have negative coordinates as they do not belong to the non-negative orthant. Further, $[f_{\text{c}}\leq t]^+$ has extra extreme points that result from the intersection of $f_{\text{c}}\leq t$ and planes $y_s^n=0$. Norm-ball $f_{\text{c}}$ is symmetric w.r.t. every plane $y^n_s=0$ and hence the extreme point resulting from the intersection of the norm-ball and such a plane will either be an extreme point of the norm ball or the midpoint of two extreme points of the norm-ball with $y^n_s$ coordinates of $+1$ and $-1$. In the latter case, the $y^n_s$ coordinate of the extreme point of $[f_{\text{c}}\leq t]^+$ will be 0. Either case, $\text{Supp}(\ve{v})$ will still be a  subset of $[N]_{(s-1)N}$ for an $s\in[K-1]$. If there is no nonzero entry left in the coordinates of the extreme point of the intersection, which is the case when all planes $y^n_s=0$ intersect, the resulting point is the origin.
\end{IEEEproof}

We now prove Thereom~\ref{th:optimal}.
\begin{IEEEproof}[Proof of Theorem~\ref{th:optimal}]
At optimality, we have $f_{\text{c}}(\tilde{\ve{y}}^*)=t^*$, as otherwise the objective can be decreased by replacing $t^*$ with $f_{\text{c}}(\tilde{\ve{y}}^*)$, which contradicts the optimality of $t^*$. 

The objective function (\ref{eq:objective_sumodular}) calculated at an extreme point of form  (\ref{eq:extPointsFeas}) with nonzero parameters for $s=s_o$ and $A$ is $[1+\frac{\sum_{n\in g^{-1}(A)}(s_o/K-1+\alpha^*_n)\gamma-Kp_n\alpha^*_n}{(K-s_o)/(s_o+1)(1-(1-P(A))^{s_o+1}}]t$, which is a factor of $t$. Denote the denominator of (\ref{eq:extPointsFeas}) by $t^u(s,A)$, i.e., $t^u(s,A)=\frac{K-s}{s+1}[1-(1-P(A))^{s+1}]$. Notice that for $t=t^u(s_o,A)$, the extreme points of $[f_{\text{c}}\leq t]^+$ are of form $y^n_s=1$ for $s=s_o, n\in g^{-1}(A)$, and $y^n_s=0$ otherwise. These parameters satisfy (\ref{eq:ballConstraint_submodular})-(\ref{eq:boxConstraint_submodular}) and are feasible. Hence, for any $s_o\in [K-1]$ and $A\subset [N]_{(s_o-1)N}$, we have
\begin{align*}
\left[1\!+\!\frac{\sum_{n\in g^{-1}(A)}(\frac{s_o}{K}-1+\alpha^*_n)\gamma-Kp_n\alpha^*_n}{\frac{K-s_o}{s_o+1}(1-(1-P(A))^{s_o+1}}
\right]\!t^u(s_o,A)
&\geq t^*\!+\!\sum_{n=1}^N\sum_{s=1}^{K-1}[(\frac{s}{K}\!-\!1\!+\!\alpha^*_n)\gamma-Kp_n\alpha_n^*](y^n_s)^{*}\\
&=\!\left[\!1\!+\!\frac{\sum_{n\in g^{-1}(A^*)}(\frac{s^*}{K}\!-\!1\!+\!\alpha^*_n)\gamma\!-\!Kp_n\alpha^*_n}{\frac{K-s^*}{s^*+1}(1-(1-P(A^*))^{s^*+1}}\right]\!\!t^*
\end{align*}
where the equality holds as the extreme points of $[f_{\text{c}}\leq t^*]^+$  are in the form of (\ref{eq:extPointsFeas}), and one of them, say $\bar{\ve{y}}$, with $s=s^*$ and $A=A^*$ has the smallest objective (\ref{eq:objective_sumodular}) among the extreme points. Since the inequality holds for every $s_o\in [K-1]$ and $A\subset [N]_{(s_o-1)N}$, it also holds for $s=s^*$ and $A=A^*$ in the LHS. This yields $t^u(s^*,A^*)\geq t^*$ and equivalently $\frac{t^*}{t^u(s^*,A^*)}\leq 1$. As a result, the extreme point $\bar{\ve{y}}$ also satisfies (\ref{eq:partitionConstraint_submodular}) and is feasible to (\ref{eq:optimProblem_submodular}). Given that it has the smallest objective among the extreme points of $[f_{\text{c}}\leq t^*]^+$, it is optimal, i.e., $\bar{\ve{y}}=\ve{y}^*$.

Now, the objective $\left[1+\frac{\sum_{n\in g^{-1}(A^*)}(\frac{s^*}{K}-1+\alpha^*_n)\gamma-Kp_n\alpha^*_n}{\frac{K-s^*}{s^*+1}(1-(1-P(A^*))^{s^*+1}}\right]t^*$ is linear in $t^*$. Since $t^*\leq t^u(s^*,A^*)$ and $t^*=t^u(s^*,A^*)$ is achievable, at optimality we either have $t^*=0$ or $t^*=t^u(s^*,A^*)$, depending on the sign of the coefficient of $t^*$. In the former case, we either have cached all files, i.e., $\forall n: (y^n_K)^{*}=1,(y^n_s)^{*}=0, s<K$, or no file is cached at all, i.e., $\forall n: (y^n_0)^{*}=1,(y^n_s)^{*}=0, s>1$, as in both cases the rate $f_{\text{c}}$ due to delivery of the content cached in at least one cache is 0. 

In the case of  $t^*=t^u(s^*,A^*)$, for $s\in [K-1]$ we have $(y^n_s)^{*}=1, s=s^*, n\in g^{-1}(A^*)$ and $(y^n_s)^{*}=0$ otherwise. Together with Lemma~\ref{lem:binary}, this concludes that at optimality $(z^n)^*=1-\sum_{s=1}^{K-1} (y^n_s)^{*}\in\{0,1\}$. Hence, when $(z^n)^{*}=1$ we have $(y^n_0)^{*}=1$ and $(y^n_K)^{*}=0$ if $Kp_n< \gamma$ and $(y^n_0)^{*}=0$ and $(y^n_K)^{*}=1$ if $Kp_n\geq \gamma$.
\end{IEEEproof}
\section{Proof of Lemma~\ref{lem:gamma_M}}\label{app:lem_gamma_M}
To prove Lemma~\ref{lem:gamma_M}, we first show the following result:
\begin{lemma}\label{lem:equality}
The capacity constraint (\ref{eq:capacityConstraint_rmsc_full}) in \ac{rmsc} is  satisfied with equality at optimality, i.e., no storage remains unused.
\end{lemma}
\begin{IEEEproof}
Assume that for storage capacity $M$, there is an optimal solution $\ve{y}^*$ with $m(\ve{y}^*)+\epsilon N=M$, where $\epsilon>0$. Then, construct solution $\ve{y}'$ with $y'^n_s=(1-\epsilon)y^n_s,s<K$ and $y'^n_K=(1-\epsilon)y^n_K+\epsilon$. Essentially, $\ve{y}'$ splits every file into two parts of lengths $(1-\epsilon)F$ and $\epsilon F$. It uses $\ve{y}^*$ for the placement of the parts of length $(1-\epsilon)F$ and caches the other $\epsilon F$ parts on every cache. This uses $(1-\epsilon)m(\ve{y}^*)+\epsilon\leq (1-\epsilon)M+\epsilon N< M$ of storage, which implies that the storage constraint is satisfied for $\ve{y}'$. However, $r(\ve{y}')=(1-\epsilon)r(\ve{y}^*)+\epsilon\times 0< r(\ve{y}^*)$. This contradicts the optimality of $\ve{y}^*$.
Hence, the optimal solution of \ac{rmsc} must satisfy the capacity constraint by equality.
\end{IEEEproof}
\begin{IEEEproof}[Proof of Lemma~\ref{lem:gamma_M}]
The first property follows from the shadow price interpretation of the Lagrange multipliers for inequality constraints \cite[Section~5.6]{Boyd:2004}. In particular, let denote the optimal solutions to (\ref{eq:optimProblem_rmsc_full}) with storage budgets $M_1$ and $M_2<M_1$ by $\ve{y}^*_1$ and $\ve{y}^*_2$. Then, $r(\ve{y}^*_1)\leq r(\ve{y}^*_2)$. Since duality gap is zero, the primal and dual objectives are equal, this implies $r(\ve{y}^*_1)+\gamma_1^*m(\ve{y}_1^*)\leq r(\ve{y}^*_2)+\gamma_2^*m(\ve{y}_2^*)$. Hence, $\gamma^*_1\leq \gamma^*_2$ as otherwise $ r(\ve{y}^*_1)+\gamma_2^*m(\ve{y}_1^*)< r(\ve{y}^*_1)+\gamma_1^*m(\ve{y}_1^*)\leq r(\ve{y}^*_2)+\gamma_2^*m(\ve{y}_2^*)$, which contradicts optimality of $\ve{y}_2^*$.

The second property follows from the fact that the set $\gamma\geq 0$ in the Lagrangian minimization problem (\ref{eq:dualFun}), or equivalently in \ac{jrsm}, is continuous, while set $\mathcal{Y}^*$ (or $\mathcal{M}$) is finite. Hence, a range of values of $\gamma$ must map to the same storage $M\in\mathcal{M}$ and they are all dual optimal. 

To prove the third property consider $\ve{y}_1,\ve{y}_2\in\mathcal{Y}^*$ that correspond to two consecutive storage values $M_1=m(\ve{y}_1)$ and $M_2=m(\ve{y}_2)$. Without loss of generality assume that $M_1<M_2$. Clearly, $M\not\in \mathcal{M}$ for any $M\in(M_1,M_2)$. Now, notice that i) each capacity $M$ must correspond to some $\gamma^*$ in the dual problem, ii) for each $\gamma\geq 0$ there is an optimal solution to \ac{jrsm} in $\mathcal{Y}^*$ and a corresponding storage value in $\mathcal{M}$, hence none of those solutions uses an amount of storage $M\not\in\mathcal{M}$ and  iii)  based on Lemma~\ref{lem:equality}, at optimality all the available storage must be used, and iv) based on property 1, $\gamma^*$ is nondecreasing in $M$. These point conclude that the optimal dual parameter for any $M\in[M_1,M_2]$ must belong to $ \{\gamma^*_{M_1}, \gamma^*_{M_2}\}$, where $\gamma^*_{m}$ represents the optimal dual parameter for capacity $m$ and based on property 1, $\gamma^*_{M_2}\leq \gamma^*_{M_1}$. More specifically, point (iv) requires $\gamma^*_{M}=\gamma^*_{M_1}$ for $M\in [M_1,M']$ and $\gamma^*_M=\gamma^*_{M_2}$ for $M\in (M',M_2]$, for some $M_1\leq M'\leq M_2$. However, we must have $\gamma^*_{M_2}= \gamma^*_{M_1}$ as otherwise for any $\gamma^*_{M_2}<\gamma''< \gamma^*_{M_1}$ corresponds to a value of $M\not\in\mathcal{M}$, which contradicts point (ii). This concludes property 3, i.e., for two consecutive values $M_1,M_2\in\mathcal{M}, M_1<M_2$, all capacities $M_1\leq M
\leq M_2$ correspond to the same dual parameter $\gamma^*$.
Further, we can derive the optimal dual parameter for 
$m(\ve{y}_1)\leq M\leq\tilde{m}(\ve{y}_2)$ as it satisfies 
$r(\ve{y}_1)+\gamma^* m(\ve{y}_1)=r(\ve{y}_2)+\gamma^* m(\ve{y}_2)=L^*$. Hence,
\begin{align}\label{eq:gamma_two}
\gamma^*=\frac{r(\ve{y}_1)-r(\ve{y}_2)}{m(\ve{y}_2)-m(\ve{y}_1)}.
\end{align}
\end{IEEEproof}
\section{Proof of Theorem~\ref{th:optimal_rmsc}}\label{app:th2_proof}
\begin{IEEEproof} To prove Theorem~\ref{th:optimal_rmsc}, we consider two cases:
\paragraph*{Case I ($M\in\mathcal{M}$)} This case is straightforward because of the zero duality gap in the primal-dual framework established in Section~\ref{sec:pr_du}. In particular, vector $\ve{y}_{\text{\small\ac{jrsm}}}^*\in\mathcal{Y}^*$ with $m(\ve{\ve{y}_{\text{\small\ac{jrsm}}}^*})=M$ is also optimal to \ac{rmsc}.
\footnote{A direct proof for optimality of $\ve{y}^*_{\tiny\ac{jrsm}}$ with $m(\ve{y}^*_{\tiny\ac{jrsm}})=M$ for \ac{rmsc} is as follows. Based on Lemma~\ref{lem:equality}, the optimal solution of \ac{rmsc} satisfies the capacity constraint with equality. Now, assume that $\ve{y}^*_{\tiny\ac{jrsm}}$ is not optimal for the \ac{rmsc} problem. This means that $r(\ve{y}^*_{\tiny\ac{rmsc}})<r(\ve{y}^*_{\tiny\ac{jrsm}})$. However, since $m(\ve{y}^*_{\tiny\ac{rmsc}})=m(\ve{y}^*_{\tiny\ac{jrsm}})=M$, this implies that $r(\ve{y}^*_{\tiny\ac{rmsc}})+\gamma^*m(\ve{y}^*_{\tiny\ac{rmsc}})<r(\ve{y}^*_{\tiny\ac{jrsm}})+\gamma^*m(\ve{y}^*_{\tiny\ac{jrsm}})$. The last result contradicts the optimality of $\ve{y}^*_{\tiny\ac{jrsm}}$ for \ac{jrsm}. Hence, $\ve{y}^*_{\tiny\ac{jrsm}}$ must be optimal for the \ac{rmsc} problem.}

\paragraph*{Case II ($M\not\in\mathcal{M}$)} 
To derive the optimal solution of \ac{rmsc} for $M\not\in\mathcal{M}$, we use Lemma~\ref{lem:gamma_M}.
Let $\ve{y}_1,\ve{y}_2\in\mathcal{Y}^*$ be the solutions corresponding to the two consecutive storage values $m(\ve{y}_1),m(\ve{y}_2)\in \mathcal{M}$ such that $m(\ve{y}_1)< M<m(\ve{y}_2)$. Let $\gamma^*$ and $L^*$ be the corresponding optimal dual parameter and Lagrangian value, respectively. 
Since $m(\cdot)$ is linear, for any given storage $m(\ve{y}_1)< M<m(\ve{y}_2)$, there exists a convex combination of $\ve{y}_1$ and $\ve{y}_2$  that uses storage $M$. We show that the same convex combination also minimizes the Lagrangian for $\gamma^*$. In that case, we are back to a case similar to Case I, and the same argument used there requires the convex combination to also optimize \ac{rmsc}.

Consider $\ve{y}_{\theta;\ve{y}_1,\ve{y}_2}\delequal\theta \ve{y}_1 + (1-\theta)\ve{y}_2$ for $0<\theta<1$. 
For the $\theta$ for which $m(\ve{y}_{\theta;\ve{y}_1,\ve{y}_2})=M$,  we need to show that $\ve{y}_{\theta;\ve{y}_1,\ve{y}_2}$ minimizes the Lagrangian for $\gamma^*$. This is equivalent to showing that $r(\ve{y}_{\theta;\ve{y}_1,\ve{y}_2})+\gamma^*m(\ve{y}_{\theta;\ve{y}_1,\ve{y}_2})=L^*$.\footnote{The equivalence results from the fact that if $r(\ve{y}_{\theta;\ve{y}_1,\ve{y}_2})+\gamma^*m(\ve{y}_{\theta;\ve{y}_1,\ve{y}_2})<L^*$, then $\ve{y}_1$ and $\ve{y}_2$ could not be optimal for $\gamma^*$, which is a contradiction.}  Since $m(\ve{y}_{\theta;\ve{y}_1,\ve{y}_2})=\theta m(\ve{y}_1) + (1-\theta)m(\ve{y}_2)$, it is sufficient to that  $r(\ve{y}_{\theta;\ve{y}_1,\ve{y}_2})=\theta r(\ve{y}_1) + (1-\theta)r(\ve{y}_2)$  to prove $r(\ve{y}_{\theta;\ve{y}_1,\ve{y}_2})+\gamma^*m(\ve{y}_{\theta;\ve{y}_1,\ve{y}_2})=L^*$.
To show $r(\ve{y}_{\theta;\ve{y}_1,\ve{y}_2})=\theta r(\ve{y}_1) + (1-\theta)r(\ve{y}_2)$, notice that each $\ve{y}\in\mathcal{Y}^*$ has nonzero parameters $y^n_s$ for at most two values of $s$, one $s=0$ and one $s\geq 1$. Assume that $\ve{y}_1$ and $\ve{y}_2$ have nonzero entries respectively for $s_1>0$ and $s_2>0$ and possibly for  $s=0$. We consider two cases of $s_1\neq s_2$ and $s_1=s_2$. In the former case, $r(\ve{y}_1)=Kp_n{y_1}^n_0+\frac{K-s_1}{s_1+1}(1-(1-\sum_{i=1}^{n_1})^{s_1+1}){y_1}^n_{s_1}$, $r(\ve{y}_2)=Kp_n{y_2}^n_0+\frac{K-s_1}{s_1+1}(1-(1-\sum_{i=1}^{n_2})^{s_1+1}){y_2}^n_{s_1}$ and $r(\theta \ve{y}_1 + (1-\theta)\ve{y}_2))=Kp_n(\theta{y_1}^n_0+(1-\theta)\theta{y_2}^n_0)+\theta\frac{K-s_1}{s_1+1}(1-(1-\sum_{i=1}^{n_1})^{s_1+1}){y_2}^n_{s_1}+(1-\theta)\frac{K-s_2}{s_2+1}(1-(1-\sum_{i=1}^{n_2})^{s_2+1}){y_2}^n_{s_2}=\theta\tilde{r}(\ve{y}_1)+(1-\theta)r(\ve{y}_2)$. For the case of $s_1=s_2=s_o$, let $n_1$ and $n_2$ be the smallest indexes $n$ with nonzero $y^n_{s_o}$ in the corresponding two solutions. Then, since $m(\ve{y}_1)<m(\ve{y}_2)$, we must have $n_2>n_1$. Hence, for the rates we have
\begin{align*}
r(\ve{y}_1)&= kp_n{y^n_0}_1 +\frac{K-s_o}{s_o+1}\sum_{g\in\mathcal{G}_{s_o+1}} \pi_{s_o+1}^g \max_{n\in g}{y^n_{s_o}}_1=kp_n{y^n_0}_1 +\frac{K-s_o}{s_o+1}\sum_{g\in\mathcal{G}_{s_o+1},g\cap[n_1]\neq \emptyset} \pi_{s_o+1}^g \\
r(\ve{y}_2)&= kp_n{y^n_0}_2 +\frac{K-s_o}{s_o+1}\sum_{g\in\mathcal{G}_{s_o+1}} \pi_{s_o+1}^g \max_{n\in g}{y^n_{s_o}}_2=kp_n{y^n_0}_2 +\frac{K-s_o}{s_o+1}\sum_{g\in\mathcal{G}_{s_o+1},g\cap[n_2]\neq \emptyset} \pi_{s_o+1}^g
\end{align*}
and
\begin{align*}
r(\theta \ve{y}_1 + (1-\theta)\ve{y}_2)&= kp_n(\theta {y^n_0}_1 + (1-\theta){y^n_0}_2) +\frac{K-s_o}{s_o+1}\sum_{g\in\mathcal{G}_{s_o+1}} \pi_{s_o+1}^g \max_{n\in g}\theta {y^n_{s_o}}_1 + (1-\theta){y^n_{s_o}}_2\\
&=kp_n(\theta {y^n_0}_1 + (1-\theta){y^n_0}_2)+\frac{K-s_o}{s_o+1}\sum_{g\in\mathcal{G}_{s_o+1},g\cap[n_1]\neq \emptyset} \pi_{s_o+1}^g (\theta+(1-\theta))\\
&\quad+\frac{K-s_o}{s_o+1}\sum_{g\in\mathcal{G}_{s_o+1},g\cap[n_1]=\emptyset,g\cap[n_2-n_1]_{n_1}\neq\emptyset} \pi_{s_o+1}^g (1-\theta)
\\
&=\theta\left[kp_n{y_1}^n_0 +\frac{K-s_o}{s_o+1}\sum_{g\in\mathcal{G}_{s_o+1},g\cap[n_1]\neq \emptyset=\emptyset} \pi_{s_o+1}^g\right]\\
&\quad+(1-\theta)\left[kp_n{y_2}^n_0+\frac{K-s_o}{s_o+1}\sum_{g\in\mathcal{G}_{s_o+1}\emptyset,g\cap[n_2]\neq\emptyset} \pi_{s_o+1}^g \right]\\
&=\theta\tilde{r}(\ve{y}_1)+(1-\theta)r(\ve{y}_2)
\end{align*}
where we used the fact that if $g\cap[n_1]\neq \emptyset$, then $n_2>n_1$ implies $g\cap[n_2]\neq\emptyset$. This completes the proof of the third feature as we now have
\begin{align*}
r(\ve{y}_{\theta;\ve{y}_1,\ve{y}_2})+\gamma^*m(\ve{y}_{\theta;\ve{y}_1,\ve{y}_2})&=\theta\tilde{r}(\ve{y}_1)+(1-\theta)r(\ve{y}_2)+\gamma^*[\theta\tilde{m}(\ve{y}_1)+(1-\theta)m(\ve{y}_2)]\\
&=\theta[r(\ve{y}_1)+\gamma^*m(\ve{y}_1)]+(1-\theta)[r(\ve{y}_2)+\gamma^*m(\ve{y}_2)]\\
&=\theta L^*+(1-\theta)L^*=L^*.
\end{align*}
\end{IEEEproof}

\bibliographystyle{IEEEtran}
\bibliography{CachingJune16.bib}

\end{document}